\documentclass[aps,prd,twocolumn,showkeys,superscriptaddress]{revtex4-2}
\usepackage{amsmath,amssymb}
\usepackage{graphicx}

\newcommand{\bastar}{\begin{eqnarray*}}
\newcommand{\eastar}{\end{eqnarray*}}
\newskip\humongous \humongous=0pt plus 1000pt minus 1000pt

\newif\ifdtup

\relax
\newcommand{\hr}{{\hat r}}

\newcommand{\bea}{\begin{eqnarray}}
\newcommand{\eea}{\end{eqnarray}}
\newcommand{\pd}{\partial}
\newcommand{\A}{{\vec A}}
\newcommand{\F}{{\vec F}}

\newcommand{\nn}{\nonumber}

\newcommand{\Int}{\displaystyle{\int}}
\newcommand{\mn}{{\mu\nu}}

\begin{document}
\title{Electroweak Primordial Magnetic Blackhole: Cosmic Production and Physical Implication}
\bigskip
\author{Y. M. Cho}
\email{ymcho0416@gmail.com}
\affiliation{School of Physics and Astronomy, 
Seoul National University, Seoul 08826, Korea}
\affiliation{Center for Quantum Spacetime, Sogang University,
Seoul 04107, Korea}  
\author{Sang-Woo Kim}
\email{sangwoo7616@gmail.com}
\affiliation{Department of Liberal Arts,
Konkkuk University, Seoul 143701, Korea}
\author{Seung Hun Oh}
\email{shoh.physics@gmail.com}
\affiliation{Department of Liberal Arts,
Korea Polytechnic University,
Siheung 15073, Korea}

\begin{abstract}
The electroweak monopole, when coupled to gravity, 
turns to the Reissner-Nordstrom type primordial magnetic blackhole whose mass is bounded below, with the lower bound $M_P \sqrt \alpha$. This changes 
the overall picture of the monopole production mechanism in the early universe drastically and has deep implications in cosmolpgy. In particular, this enhances the possibility that the electroweak monopoles (turned to the primordial magnetic blackholes) could become the seed of stellar objects and galaxies, and account for the dark matter 
of the universe. Moreover, this tells that we have
a new type of primordial blackhole different from 
the popular primordial blackhole in cosmology, 
the electroweak primordial magnetic blackhole based 
on a totally different production mechanism. We discuss the physical implications of the electroweak primordial magnetic blackhole.
\end{abstract}


\maketitle

\section{Introduction}

Since Dirac proposed the magnetic monopole generalizing 
the Maxwell's theory, the monopole has become an obsession 
in physics, experimentally as well as theoretically \cite{dirac,cab}. After the Dirac monopole we have had 
the Wu-Yang monopole, the 't Hooft-Polyakov monopole, 
and the grand unification monopole \cite{wu,thooft,dokos}. 
But the electroweak monopole (also known as the Cho-Maison monopole) stands out as the most realistic monopole 
that could actually exist in nature and be detected \cite{plb97,yang,epjc15,ellis,bb,epjc20,bbc}. This is 
because it exists in the standard model. So, if 
the standard model is correct as we believe, this monopole must exist.

Indeed the Dirac monopole in electrodynamics should 
transform to the electroweak monopole after 
the unification of the electromagnetic and weak 
interactions, and the Wu-Yang monopole in QCD is 
supposed to make the monopole condensation to confine 
the color. Moreover, the 'tHooft-Polyakov monopole 
exists only in a hypothetical theory, and the grand unification monopole which could have been amply produced 
at the grand unification scale in the early universe 
probably has become completely irrelevant at present 
universe after the inflation. 

This makes the experimental confirmation of the electroweak 
monopole one of the most urgent issues in the standard 
model after the discovery of the Higgs particle. In fact 
the detection of this monopole, not the Higgs particle, 
should be regarded as the final test of the standard 
model. For this reason the newest MoEDAL detector at 
LHC is actively searching for the monopole \cite{medal1,medal2,medal3,atlas}. 

To detect at LHC, we need to keep in mind some basic 
facts about the monopole \cite{plb97}. First, the magnetic charge of the electroweak monopole is not $2\pi/e$ but $4\pi/e$, twice that of the Dirac monopole. This is 
because the period of the electromagnetic U(1) subgroup 
of the standard model becomes $4\pi$, not $2\pi$. 
This comes from the fact the electromagnetic U(1)  
comes (partly) from the U(1) subgroup of SU(2). Second, 
the mass of the monopole is estimated to be of the order 
of several TeV, or roughly $M_W/\alpha$. This is because the mass basically comes from the same Higgs mechanism which makes the W boson massive, except that here the coupling is magnetic (i.e., $4\pi/e$). This makes 
the monopole mass $1/\alpha$ times heavier than the W 
boson mass, of the order of 11 TeV. Third, in spite of this, the size of the monopole is set by the W boson mass. This is because the monopole solution has the weak boson dressing which shows that the size is fixed by the W 
boson masses. Finally, it exists within (not beyond) 
the standard model as the electroweak generalization 
of the Dirac monopole, which can be viewed as a hybrid between Dirac and 't Hooft-Polyakov monopoles. 

Because of these unique characteristics of the electroweak monopole MoEDAL could detect the monopole without much difficulty, if LHC could produce it. However, the 14 TeV LHC may have no chance to produce the monopole if 
the mass becomes larger than 7 TeV \cite{bbc}. In this 
case we must try to detect the remnant monopoles 
at present universe produced in the early universe.   

To detect the remnant monopoles, we need to know how 
the monopoles are produced in the early universe and 
how many of them are left over in the present universe. 
In the literature there have been discussions on 
the cosmological production of monopoles, but most 
of them have been on the grand unification 
monopole \cite{kibb,pres,guth,zurek}. The general 
consensus is that the grand unification monopoles would 
have overclosed the universe without the inflation, 
but the inflation might have completely diluted them 
in such a way that they could have no visible impact 
on the present universe \cite{infl}.

For the electroweak monopole, we have a totally different situation. A recent study showed that the electroweak 
monopole amply produced during the electroweak phase transition does not alter the standard cosmology in any significant way, but could play important roles in 
cosmology \cite{pta19}. They  could become the seed of 
the stellar objects and galaxies and thus play 
an important role in the formation of the large scale structure of the universe, and could even account for 
the dark matter of the universe. This was based on 
the observation that, as the only absolutely stable topological elementary particle which has a huge mass 
in the early universe, it could generate the density perturbation and evolve to the primordial magnetic 
blackholes (PMBHs). Moreover, the electroweak monopoles 
could generate the intergalactic magnetic field, and 
could become the source of the ultra high energy cosmic rays \cite{plb24}. 

Actually, the electroweak monopoles can turn to 
the primordial magnetic blackholes without any 
density perturbation. It has been well known that 
the monopoles, when coupled to gravity, automatically become the Reissner-Nordstrom (RN) type magnetic 
blackholes \cite{bais,prd75,plb16,yangb,wong}. 
This means that the electroweak monopoles produced 
in the early universe automatically become the PMBHs. 
This makes the above proposal more credible. On the other hand, this necessitates us to re-analyse the electroweak monopole production in the early universe more carefully.  

{\it The purpose of this paper is two-fold. The first 
purpose is to study the electroweak monopole production 
in the early universe in more detail and discuss how 
the electroweak monopoles turn to the PMBHs when 
coupled to gravity. The second purpose is to discuss physical implications of the PMBHs in cosmonogy. We 
show that the electroweak PMBHs enhance the possibility that they become the seed of the stellar objects and 
galaxies and thus play important roles in the formation 
of the large scale structure of the universe. Moreover, they enhance the possibility for the magnetic blackholes 
to become the dark matter of the universe greatly.}  

In doing so we also discuss serious shortcomings of 
the popular monopole production mechanism in the early universe. It has been generally believed that the monopole production in the early universe takes place during 
the phase transition, and that the monopole production mechanism critically depends on the type of the phase transition. In the first order phase transition 
the monopole production is thought to take place by 
the vacuum bubble collision, which makes the production probability exponentially decreasing. But in the second 
order phase transition the monopoles are supposed to be produced by the Kibble-Zurek mechanism without 
the exponential suppression. This view, however, neglects 
the fact that the monopoles are produced by the change 
of topology made by the thermal fluctuation of the Higgs vacuum, which starts at the phase transition but continues 
till the universe cools down to the Ginzburg temperature 
even after the phase transition.  

This means that the monopoles are produced not just 
during the phase transition but for quite a long time 
after the phase transition, as long as the Ginzburg temperature is considerably lower than the critical temperature. This necessitates a serious modification 
of the monopole production mechanism in the early 
universe. In fact this tells that what is important 
for the monopole production in the early universe is 
not the type of the phase transition but the Ginzburg temperature. 

Moreover, it has generally been believed that in 
the second order phase transition the initial monopole density is determined by the correlation length set 
by the Higgs mass. We find that this view also has 
a problem. This is because the monopole mass is roughly $1/\alpha$ times bigger than the W boson mass, while 
the monopole size is fixed by the W boson mass. This implies that the correct correlation length for 
the monopole production should be more likely $1/\alpha^{1/3}$ times the correlation length set 
by the Higgs mass. This could reduce the initial 
monopole density by the factor $\alpha$.    

But a most important shortcoming is that the monopole 
production in the early universe has completely 
neglected the fact that, when coupled to gravity, 
the monopole automatically transforms to magnetically charged RN type blackhole. This means that the monopoles produced in the early universe are destined to play important roles as PMBHs, so that we can not separate 
the monopoles and PMBHs in the early universe when 
we have the monopoles. And this point has not been 
properly appreciated so far.
     
To understand this, remember that the monopole mass is 
expected to be of $M_W/\alpha$, or of 11 TeV. But after 
it turns to the PMBH the mass can be anywhere between $10^{20}~\text{GeV}$ for the extremal blackholes and 
the infinity in principle for non-extremal ones. This 
means that the monopole mass changes at least by 
the factor $10^{16}$ when they become the blackholes. 
Moreover, the PMBHs and monopoles have totally different cosmic evolution. Most of the monopoles 
and antimonopoles produced around $T_i\simeq 102~{\rm GeV}$ are immediately annihilated, and this annihilation continues long time till the temperature reaches $T_f \simeq 65~{\rm MeV}$. 

But for the PMBHs this annihilation do not take place, 
so that they evolve adiabatically with the Hubble expansion. Moreover, they can not evaporate completely, because they carry the conserved magnetic charge. This means that they have a totally different impacts in cosmology. In particular, the PMBHs increase the possibility for the monopole to become the seed of the stellar objects and galaxies, and 
the dark matter of the universe greately. 

It must be emphasized that this PMBH is different from 
the well known primordial blackhole (PBH) proposed by Zeldivich and Novikov, which is supposed to be produced 
by the statistical density perturbation in the very early universe close to the Planck time \cite{zel,hawk,carr}. 
Recently the PBH has become a fascinating subject in cosmology because it could account for the dark matter 
of the universe \cite{carrk,green,carrc}. Moreover, 
it has been asserted that some of these PBHs could 
also carry color charge \cite{ak}. But it is not clear 
that the PBH proposed by Zeldovich and Novikov could account for the dark matter \cite{pbh}.

In contrast, our PMBH does not come from the density perturbation, but comes from the gravitational 
interaction of the electroweak monopole in the standard model during the electroweak phase transition. So 
the production mechanism and the production time are totally different. And we can not neglect the electroweak monopole as far as the standard model is correct. 
More importantly, our PMBH could easily account for 
the dark matter as we will see in the following. And 
some of our PMBHs could also carry color charge, because they were formed before the QCD color confinement set in.   

The paper is organized as follows. In Section II we 
briefly review the electroweak phase transition which produces the monopole. In Section III we review 
the cosmic production of the electroweak monopole. In Section IV we discuss the evolution of the electroweak monopoles and the remnant monopole density. In section 
V we discuss how the Cho-Maison monopole turns to 
the PMBH when coupled to gravity in the early universe. 
In Section VI we discuss cosmic production and 
evolution of the PMBHs. In Section VII we discuss 
the density of the remnant electroweak PMBHs at present universe and study the plausibility that they become 
the seed of the stellar objects and galaxies, and 
account for the dark matter. Finally in Section VIII 
we discuss the physical implications of the electroweak PMBH. 

\section{Electroweak Phase Transition and Ginzburg Temperature: A Review} 

\begin{figure}
\includegraphics[height=4.5cm, width=8cm]{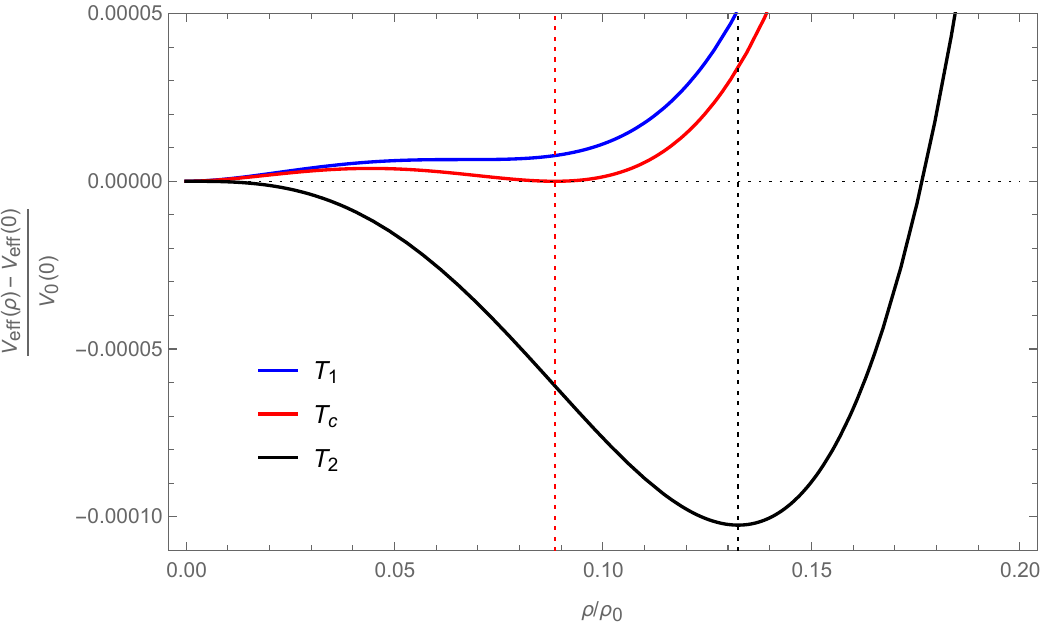}
\caption{\label{tpot} The effective potential (\ref{epot})
at $T_1,~T_c,~T_2$. Here the unit of $V_{eff}$ is chosen 
to be $V_0=(\lambda/8) \rho_0^4=1$.}
\end{figure}

To discuss the monopole production in the early universe, we start from the temperature dependent effective action 
of the standard model which describes the electroweak 
phase transition \cite{pta19,kriz,and,koba}
\begin{gather}
V_{eff}(\rho) =V_0(\rho) -\frac{C_1}{12\pi} \rho^3~T
+\frac{C_2}{2} \rho^2~T^2 
-\frac{\pi^2}{90} g_*(T) T^4,  \nn\\
V_0(\rho)=\frac{\lambda}{8}(\rho^2-\rho_0^2)^2 ,  \nn\\
C_1=\frac{6 M_W^3 + 3 M_Z^3}{\rho_0^3}\simeq 0.36,   \nn\\
C_2=\frac{4M_W^2 +2 M_Z^2 +M_H^2+4m_t^2}{8\rho_0^2} 
\simeq 0.36,   \nn\\
g_*(T) = \Sigma_B~g_B(T) + \Sigma_F~\frac{7}{8} g_F(T),
\label{epot}
\end{gather}
where $V_0$ (with $\lambda \simeq 0.26$ and 
$\rho_0 \simeq 254.6~{\rm GeV}$) is the zero-temperature potential, $C_1$ and $C_2$ terms are the loop 
contributions from the gauge bosons, Higgs field, 
and fermions, $g_*(T)$ is the total number of distinct helicity states of the particles with mass smaller than $T$, $M_W$, $M_Z$, $M_H$, and $m_t$ are the W-boson, Z-boson, Higgs boson, and the top quark masses. 
The microscopic view of the effective potential (\ref{epot}) near the critical tempertature is shown 
in Fig. \ref{tpot}.

It has three local extrema at 
\begin{gather}
\rho_s=0,   \nn\\
\rho_{\pm}(T)=\Big[\frac{C_1}{4\pi \lambda}
\pm \sqrt{ \Big(\frac{C_1}{4\pi \lambda} \Big)^2
+\Big(\frac{\rho_0}{T} \Big)^2 
-\frac{2C_2}{\lambda}} \Big]~T.
\label{rext}
\end{gather}
The $\rho_s=0$ represents the Higgs vacuum of the symmetric (unbroken) phase, the $\rho_-(T)$ represents the local maximum, and the $\rho_+(T)$ represents the local minimum Higgs vacuum of the broken phase. But the two extrema $\rho_{\pm}$ appear only when $T$ becomes smaller than 
$T_1$
\begin{gather}
T_1 =\frac{4 \pi \lambda}{\sqrt{32\pi^2\lambda C_2-C_1^2}}
~\rho_0 \simeq 146.13~\text{GeV}.
\end{gather}
So above this temperature only $\rho_s=0$ becomes the true 
vacuum of the effective potential, and the electroweak 
symmetry remains unbroken. 

At $T=T_1$ we have 
\begin{gather}
\rho_-=\rho_+=(C_1/4\pi \lambda)~T_1 \simeq 16.11~{\rm GeV},
\end{gather} 
but as temperature cools down below $T_1$ we have two 
local minima at $\rho_s$ and $\rho_+$ with 
$V_{eff}(0)< V_{eff}(\rho_+)$, until $T$ reaches 
the critical temperature $T_c$ where $V_{eff}(0)$ 
becomes equal to $V_{eff}(\rho_+)$,
\begin{gather}
T_c= \sqrt{\frac{18}{36\pi^2 \lambda C_2- C_1^2}} 
~\pi \lambda \rho_0  \simeq 146.09~{\rm GeV},   \nn\\
\rho_+(T_c)=\frac{C_1}{3\pi \lambda}~T_c
\simeq 21.47~{\rm GeV}.
\label{ctemp}
\end{gather}
So $\rho_s=0$ remains the minimum of the effective 
potential for $T>T_c$.

Below this critical temperature $\rho_+$ becomes the true 
minimum of the effective potential, but $\rho_s=0$ remains 
a local (unstable) minimum till the temperature reaches $T_2$. At $T=T_c$ the new vacuum bubbles start 
to nucleate at $\rho=\rho_+$, which takes over 
the unstable vacuum $\rho_s=0$ completely at $T=T_2$,
\begin{gather}
T_2=\sqrt{\frac{\lambda}{2C_2}}~\rho_0 
\simeq 145.82~\text{GeV},   \nn\\
\rho_+(T_2) = \frac{C_1}{2\pi \lambda}~T_2 
\simeq 32.15 \text{GeV}.
\end{gather} 
From this point $\rho_+$ becomes the only (true) minimum, 
which approaches to the well-known Higgs vacuum $\rho_0$ 
at zero temperature. This tells that the electroweak 
phase transition is of the first order. 

\begin{figure}
\includegraphics[height=4.5cm, width=7cm]{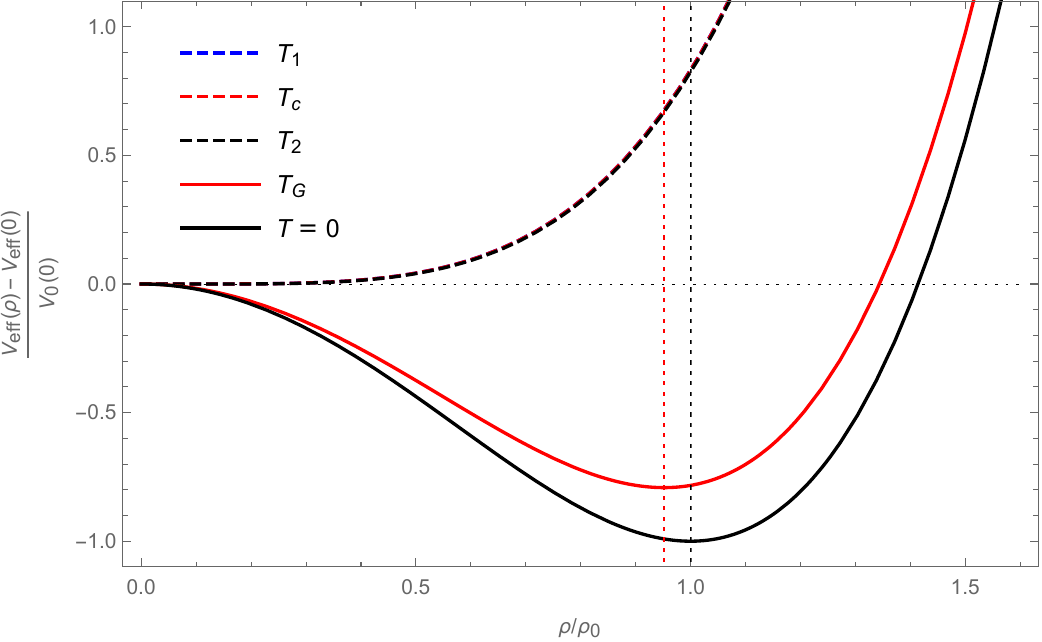}
\caption{\label{tpot1} The macroscopic view of the effective potential (\ref{epot}). The effective potentials at $T_1,~T_c,~T_2$ are shown in dotted lines, and  
the effective potentials at the Ginzburg temperature 
$T_G$ and $T=0$ are shown in red and black curves. Notice that $V_{eff}$ is almost indistinguishable at $T_1,~T_c$, 
and $T_2$.}
\end{figure}

However, the energy barrier is extremely small,
\begin{gather}
\frac{V_{eff}(\rho_-)
-V_{eff}(\rho_+)}{V_{eff}(\rho_+)}\Big|_{T_c}
\simeq 3.4 \times 10^{-6}.
\end{gather} 
Moreover, the barrier lasts only for short period since 
the temperature difference between $T_1$ and $T_c$ is 
very small, $\delta =(T_1-T_c)/T_c \simeq 0.0003$. So 
for all practical purpose we could neglect this barrier 
and treat the electroweak phase transition as a second order phase transition. The macroscopic view of 
the effective potential at different temperatures 
is shown in Fig. \ref{tpot1}, which should be compared 
with Fig. \ref{tpot}.

It has generally been believed that the monopole 
production in the early universe critically depends on 
the type of the phase transition. In the first order 
phase transition, the vacuum bubble collisions in 
the unstable vacuum are supposed to create the monopoles through the quantum tunneling to the stable vacuum 
during the phase transition, so that the monopole production is exponentially suppressed by the vacuum tunneling \cite{pres}. This, of course, is totally 
different from the monopole production mechanism in 
the second order phase transition, where the monopoles 
are created through the Kibble-Zurek mechanism without 
any exponential suppression \cite{kibb,zurek}.      

But as we have emphasized, the monopole production 
should involve a change of topology which takes place 
through the thermal fluctuation of the Higgs vacuum 
from non-vanishing to vanishing values. And this 
thermal fluctuation continues as far as \cite{gin}
\begin{gather}
\xi^3 \Delta F \le T,~~~\Delta F(T)=V(\rho_s)-V(\rho_{+}),
\label{Gcon}
\end{gather}
where $\xi(T)$ is the correlation length of the Higgs 
field given by $\xi =1/ M_H$ and $\Delta F(T)$ is 
the difference in free energy density between two 
phases. And the temperature at which this fluctuation 
stops is given by the Ginzburg temperature $T_G$,
which is defined by the condition $\xi^3 \Delta F=T$ \cite{pta19}. 

\begin{figure}
\includegraphics[height=4.5cm, width=7cm]{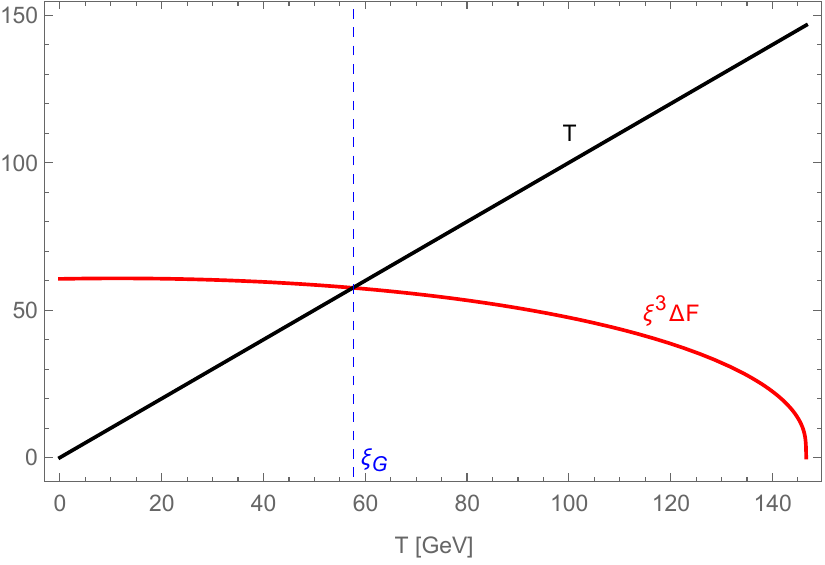}
\caption{\label{Gtemp} The determination of the Ginzburg 
temperature in the electroweak phase transition. Here 
the red and blue curve represents $\xi^3 \Delta F$ and 
the black line represents $T$.}
\end{figure}

In the electroweak phase transition we can find 
the Ginzburg temperature graphically from (\ref{epot}) 
and (\ref{Gcon}). This is shown in Fig. \ref{Gtemp}. 
From this we have \cite{pta19}
\begin{gather}
T_G \simeq  57.6~\text{GeV},
~~~~\rho_+(T_G) \simeq 232.4~\text{GeV}.
\label{gtemp}
\end{gather} 
So we can say that the monopole formation takes place between $T_2 \simeq 153.0$ GeV and $T_G \simeq 57.6$ GeV, or roughly around $T_i$,
\begin{gather}
T_i= \frac{T_c+T_G}{2} \simeq  102.09~\text{GeV},   \nn\\
\rho_+(T_i) \simeq 187.83~\text{GeV}.
\label{itemp}
\end{gather}
The effective potential (\ref{epot}) at the Ginzburg 
temperature is shown in Fig. \ref{tpot1} in red curve. 

This observation tells that the popular monopole production mechanism in the early universe has a potentially serious problem. Consider the first order phase transition. When $T_G$ becomes lower than $T_2$, we have the monopole production even after $T_2$ without any exponential suppression. In this case the monopole production in 
the first order phase transition becomes qualitatively 
the same as in the second order phase transition. 

This tells that the exponential suppression of the monopole production in the first order phase transition is only 
half of the full story which could be totally misleading. 
{\it In particular, this shows that what is important in 
the monopole production in the early universe is not 
the type of the phase transition, but the value of 
the Ginzburg temperature.} As far as the Ginzburg temperature becomes lower than $T_2$, the monopole production in the first and second order phase transitions becopmes qualitatively the same. 

\begin{figure}
\includegraphics[height=4.5cm, width=7cm]{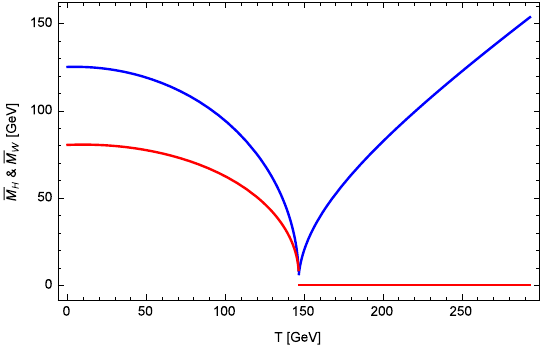}
\caption{\label{hwmass} The temperature dependent 
Higgs and W-boson masses. The blue and red curves 
represent the Higgs and W-boson masses.}
\end{figure}

The effective potential (\ref{epot}) gives us two 
important parameters of the electroweak phase transition, 
the temperature dependent Higgs mass $\bar M_H$ which determines the correlation length $\xi=1/ \bar M_H$  
\begin{gather}
\bar M_H^2=\frac{d^2 V_{eff}}{d\rho^2} 
\Big|_{\rho_{min}}  \nn\\
=\left\{\begin{array}{ll} 
\dfrac{\big[(T/T_2)^2-1 \big]}{2}~M_H^2,&~~T > T_c, \\ \dfrac{\big[(\rho_+/\rho_0)^2 +1-(T/T_2)^2 \big]}{2} 
~M_H^2, &~~T \le T_c,  	
\end{array} \right.   
\end{gather}
and the W boson mass $\bar M_W$ which determines 
the monopole mass $M_m\simeq \bar M_W / \alpha$,
\begin{gather}
{\bar M}_W^2= \left\{\begin{array}{ll} 
0,&~~~T > T_c, \\
\dfrac{g^2}{4} \rho_+^2,&~~~T \le T_c.  	
\end{array} \right. 		
\label{wbmass}
\end{gather}
The temperature dependent Higgs and W-boson masses are 
shown in Fig. \ref{hwmass}. 

Notice that $\bar M_H$ acquires its minimum value 5.5 GeV 
at $T=T_c$ and approaches to the zero temperature value 
125.2 GeV as the universe cools down. And we have
$\bar M_H (T_G) \simeq 121.7~{\rm GeV}$. On the other 
hand, the W boson starts massless at high temperature (before the symmetry breaking), but becomes massive 
toward the value 7.1 GeV at $T_c$. Moreover, we have
$\bar M_W (T_G) \simeq 76.0~{\rm GeV}$. 

This means that the monopole mass in the cosmological production is also temperature dependent. This is 
becausethe monopole mass comes from the same Higgs mechanism which makes the W boson massive. The only difference is that for the monopole, the magnetic 
coupling $4\pi/e$ makes the mass $1/\alpha$ times 
heavier. This tells that the infant monopole masses 
at $T_c$ and $T_i$ in the cosmological electroweak 
monopole production are around 0.97 TeV and 8.4 TeV, assuming $ \bar M_m  \simeq \bar M_W/\alpha$. Notice 
that $\bar M_m$ at $T_G$ already becomes almost 
the adolecent value. In Table I we show the masses 
of the Higgs, W boson and monopole at various temperatures.  

We can translate the monopole production in time scale,
converting temperature to time. In the radiation 
dominant era the age of the universe $t$ is given by \cite{kolb} 
\begin{gather}
t =\Big(\frac{90}{32 \pi G g_*(T)} \Big)^{1/2}
~\frac{1}{T^2}.
\label{ardu}
\end{gather} 
From this we have, (with $g_* \simeq 106.75$ including $\gamma, \nu, g, e, \mu, \pi, u, d, s, c, b, \tau, W, Z, H$) \cite{kolb}, 
\begin{gather}
t =0.09 \frac{M_P}{T^2}
\simeq 7 \times 10^{-7} \big(\frac{\rm GeV}{T} \big)^2 sec.	
\end{gather}  
So we can say that the electroweak monopole production 
starts from $3.3 \times 10^{-11}~sec$ to 
$2.1 \times 10^{-10}~sec$ after the big bang for 
the period of $17.7 \times 10^{-11}~sec$, or around 
$6.7 \times 10^{-11}~sec$ after the big bang in average.

\begin{table}[t]
\begin{tabular}{|c||r|r|r|r|r|}
\hline
%&&&&& \\ 
\multicolumn{1}{|c|}{} &
\multicolumn{1}{|c|}{$T$} &
\multicolumn{1}{|c|}{$\rho_{+}(T)$} &
\multicolumn{1}{|c|}{$\bar M_H(T)$} &
\multicolumn{1}{|c|}{$\bar M_W(T)$} &
\multicolumn{1}{|c|}{$\bar M_m(T)$}
\\ \hline\hline
$T_1$ & \hspace{2mm} $146.74$ & $16.3$ & $5.9$ & $0$ & $0$
\\ \hline
$T_c$ & $146.70$ & $21.7$ & $5.5$ & $7.1$ & $971.7$
\\ \hline
$T_2$ & $146.42$ & $32.5$ & $11.7$ & $10.6$ & $1\,454.8$
\\ \hline
$T_i$ & $102.09$ & $188.4$ & $92.9$ & $61.5$ & $8\,428.2$
\\ \hline
$T_G$ & $57.49$ & $232.9$ & $116.9$ & $76.0$ & $10\,419.6$
\\ \hline
$0$ & $0.00$ & \hspace{5mm} $246.2$ & \hspace{5mm} $125.3$ & \hspace{5mm} $80.4$ & $11\,014.5$
\\ \hline
\end{tabular}
\caption{The values of $\rho_{+}$, $\bar M_H$, 
$\bar M_W$, and the expected monopole mass 
$\bar M_m =\bar M_W /\alpha$ at various temperatures. 
All numbers are in GeV.}
\label{Table}
\end{table}

We can also estimate how many times the thermal 
fluctuations take place between $T_c$ and $T_G$. 
From the uncertainty principle and (\ref{itemp}) we can 
estimate the time $\Delta t$ for one fluctuation,
\begin{gather}
	\Delta t \simeq \frac{1}{\Delta E}
	\simeq 3.34 \times 10^{-27}~sec.
\end{gather} 
From this we have 
\begin{gather}
	N_f \simeq \frac{\bar t}{\Delta t}
	\simeq 3.1 \times 10^{16}.
\end{gather} 
This assures that we have (more than) enough fluctuations 
of the Higgs vacuum to produce the monopoles. 

\section{Cosmic Production of Electroweak Monopole: 
Initial Density}

The electroweak monopole production in the early universe 
and it's cosmological implications have been discussed 
before, and it has been argued that the electroweak 
monopole could have deep impact on cosmology, without 
altering the standard cosmology in any significant 
way. In particular, it has been proposed that 
the electroweak monopoles could generate the density perturbation and evolve to primordial blackholes, and 
could become the seed of the stellar objects and galaxies,  account for the dark matter of the universe, could 
generate the intergalactic magnetic field \cite{pta19}.

In this section we review the electroweak monopole production in the early universe. According to 
the Kibble's original estimate the initial electroweak monopole density is given by \cite{kibb}
\begin{gather} 
n_i \simeq \frac{g_P}{d_H^3(T_c)},
\label{Kbound}
\end{gather} 
where $d_H(T_c)$ is the horizon distance at $T_c$ and 
$g_P$ is the probability that the monopole topology is 
actually realized in one correlation volume, which has 
generally been assumed to be around 10 \% in 
the literature \cite{kriz,and}. Since $d_H=2t$ in 
the radiation dominant era we have
\begin{gather} 
d_H(T_c) =2 t(T_c) 
\simeq \frac{2.6 \times 10^{15}}{T_c}  
\simeq 0.35~cm,  \nn\\
n_i \simeq  \Big(\frac{T_c}{2.6 \times 10^{15}} \Big)^3  \times g_p \simeq 23 \times g_P~cm^{-3}.
\label{Kbound1}
\end{gather} 
As we have emphasized, however, this estimate (based on 
the assumption that the monopole is produced at $T_c$) 
is not realistic. Since the monopoles are produced not 
just during but even after the phase transition till 
the universe cools down to the Ginzburg temperature, 
we have to replace $d_H$ in (\ref{Kbound}) by the mean 
value of two correlation lengths at $T_c$ and 
$T_G$ \cite{pta19},
\begin{gather}
\xi_i =\frac{1}{\bar M_H(T_i)}
\simeq 2.1 \times 10^{-16}~\text{cm}.
\label{mcl0}
\end{gather} 
Notice that this is near the electroweak scale 
($1/M_W\simeq 2.5 \times 10^{-16}~cm$). From this we 
can estimate the initial monopole density,  
\begin{gather}
n_i \simeq \frac{g_P}{\xi_i^3}
=g_p \times \bar M_H(T_i)^3 \nn\\
\simeq  1.04 \times 10^{47} \times g_p~cm^{-3}.
\label{Cbound}
\end{gather} 
This should be compared with (\ref{Kbound1}). Obviously this huge difference comes from the difference between $d_H$ and $\xi_i$. 

However, this estimate may also have a defect. Since 
the monopole size is fixed by the electroweak scale while the mass becomes $1/\alpha$ times bigger than the W boson mass, there appears the possibility that the energy within one correlation volume may not become enough to create one monopole. So we have to make sure that the radiation 
energy in one correlation volume is no less than 
the monopole mass, and should require  
\begin{gather}
E = \rho(T_i) \times \xi_i^3 
\geq \frac{\bar M_W(T_i)}{\alpha} \simeq 8.4~\text{TeV},
\label{econ}
\end{gather}  
where $\rho(T_i)$ is the energy density of 
the universe at $T_i$. Now, with 
\begin{gather}
\rho(T_i) =\frac{\pi^2}{30}~g_*~T_i^4
\simeq 34.8~T_i^4, 
\label{red} 
\end{gather}
we can calculate the energy within one correlation 
volume at $T_i$,
\begin{gather}
E= \rho(T_i) \times \xi_i^3
\simeq \frac{\pi^2}{30}~g_*
~\big(\frac{T_i}{\bar M_H(T_i)}\big)^3~T_i \nn\\
\simeq 4.73~{\rm TeV}.
\end{gather} 
This is problematic, because this tells that 
the energy condition (\ref{econ}) is not satisfied. 

A logical way to cure this problem is to make 
the correlation length $\xi$ by $(1/\alpha)^{1/3}$
times bigger. So we introduce a new correlation 
length $\bar \xi$ by
\begin{gather}
\bar \xi_i =\big(\frac{1}{\alpha} \big)^{1/3}
\times \frac{1}{\bar M_H(T_i)}
\simeq 1.08 \times 10^{-15}~\text{cm}.
\label{mcl1}
\end{gather}  
With this we have the new initial monopole density  
\begin{gather}
n_i \rightarrow \frac{g_P}{\bar \xi_i^3}
=g_p \times \alpha~\bar M_H(T_i)^3 \nn\\
\simeq  7.6 \times 10^{44} \times g_p~cm^{-3}.
\label{Cbound1}
\end{gather} 
This should be compared with (\ref{Cbound}).

\section{Evolution of Electroweak Monopole and Remnant Monopole Density}

The initial monopole density changes as the universe  
evolves. There are two processes which changes 
the initial monopole density, the Hubble expansion 
and the annihilation of monopole-antimonopole pairs, 
and the evolution of the monopole density $n$ is 
governed by the Boltzmann equation \cite{kibb,pres}
\begin{gather}
\frac{d n}{dt} + 3 H n = -\sigma n^2.
\label{boltz1}
\end{gather}
where $H$ and $\sigma$ are the Hubble expansion 
parameter and the monopole-antimonopole annihilation 
cross section. 

\begin{figure}
\includegraphics[height=4.5cm, width=7cm]{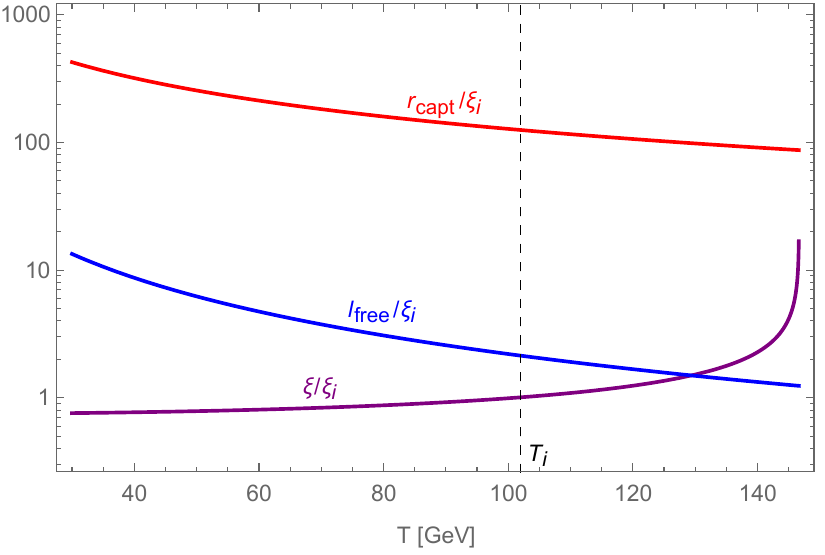}
\caption{\label{scales} The relevant scales, $\xi$ in 
purple, $l_m$ in blue, and $r_\text{capt}$ in red, 
against $T$. They are normalized by the correlation 
length $\xi_i$ at $T_i$. Here we set 
$M_m(T_i)=8.4~{\rm TeV}$.}
\end{figure}

The annihilation of the monopoles is controlled by two 
things, the thermal Brownian motion (random walk) of 
the monopole in hot thermal bath and the magnetic attraction between monopole and anti-monopole. After 
the creation the monopoles diffuse in hot thermal
plasma by the Brownian motion with the mean free path 
$l_m$ given by
\begin{gather}
l_m =v_t t_m 
\simeq \frac{1}{BT} \sqrt{\frac{ M_m}{T}},  
~~~B =\frac{1}{T}{\sum_i k_i \sigma_i},
\label{lfree}
\end{gather}
where $v_t\simeq \sqrt{T/M_m}$ and $t_m$ are the thermal velocity and the mean free time of the monopoles, 
$k_i$ and $\sigma_i$ are the number density and 
the cross section of the $i$-th relativistic charged particles and the sum is the sum over all spin 
states \cite{pres}. With 
\begin{gather}
k_i \simeq \frac{3\zeta(3)}{4\pi^2} T^3, 
~~~\sigma_i \simeq \Big(\frac{q_i}{e}\Big)^2 \frac{1}{T^2},
\end{gather}
we have 
\begin{gather}
B\simeq \frac{3\zeta(3)}{4\pi^2} 
\sum_i \Big(\frac{q_i}{e}\Big)^2 
\simeq 0.09 \times \sum_i \Big(\frac{q_i}{e}\Big)^2,
\end{gather}
where $q_i$ is the electric charge of the $i$-th particle and $\zeta$ (with $\zeta(3)=1.202...$) is 
the Riemann zeta function. Since the charged particles in the plasma are the leptons and quarks, we may put $B\simeq 3$. From this we have $v_t\simeq 0.1$ and $l_m \simeq 5.8 \times 10^{-16}~cm$ around $T_i$. 

Against the thermal random walk of the monopoles, 
the Coulombic attraction between monopole and 
anti-monopole makes them drift towards each other. 
The drift velocity $v_d$ of the monopole at a distance 
$r$ from the anti-monopole in the non-relativistic approximation is given by \cite{vil}
\begin{gather}
v_d \simeq \frac1{\alpha} \times \frac1{B T^2 r^2}.
\end{gather}
Notice that we have $v_d  \simeq 7.4$ with $r=\bar \xi_i$. 
This is unrealistic, which is due to the non-relativistic approximation. But this does tell that $v_t \ll v_d$, 
which shows that the Coulombic magnetic attraction 
is much stronger than the thermal diffusion. 
In Fig. \ref{scales} we plot the relevant scales 
$\xi$, $l_m$, and $r_\text{capt}$, against $T$ for comparison. This clearly shows that the capture radius 
is much bigger than the mean free length and 
correlation length in a wide range of $T$. This 
tells that (\ref{Cbound1}) is an overestimation.

Now, assuming the mean distance $r$ between the monopole 
and anti-monopole is $r\simeq n^{-1/3}$, we can express
the capture time by
\begin{gather}
t_\text{capt}\simeq \frac{r}{v_d} 
\simeq \alpha \times \frac{B T^2}{n}.
\end{gather}
From this we have the monopole-antimonopole annihilation cross section 
\begin{gather}
\sigma \simeq \frac{1}{t_\text{capt} n} 
= \frac{1}{\alpha B T^2}.
\end{gather}
So, with
\begin{gather}
H=\frac{\dot R}{R}= \frac{T^2}{C M_P},   
~~~C=\Big(\frac{45}{4\pi^3 g_*} \Big)^{1/2}
\simeq 3.74,	
\end{gather} 
we can express the Boltzmann equation in term of $\tau=M_m/T$ 
\begin{gather}
\frac{d}{d\tau} \Big(\frac{n_m}{T^3} \Big) 
=-\frac{CM_P}{\alpha BM_m} \Big(\frac{n_m}{T^3}\Big)^2
=-\frac{\sigma T^3}{\tau H}\Big(\frac{n_m}{T^3}\Big)^2. \label{boltz2}
\end{gather}
Solving this we have \cite{pta19}
\begin{gather}
n =\frac{T^3}{A(M_m/T - M_m/T_i) +B}, \nn\\
A= \frac{CM_P}{\alpha BM_m}, 
~~~~B=\frac{T_i^3}{n_i}, 
\end{gather}
where $M_m$ is treated as a constant. But in reality 
it depends on time, so that it should be understood 
as a mean value. 

Notice that when $\tau_i \ll \tau$, only the first term 
in the denominator becomes important. So the monopole 
density approaches to 
\begin{gather}
n \rightarrow \alpha \times \frac{BT^4}{C M_P},
\end{gather}
regardless of the initial condition \cite{pres}. 
The diffusive capture process is effective only when 
$l_m < r_{capt}$, which determines the temperature 
$T_f$ below which the monopole-antimonopole 
annihilation ceases,
\begin{gather}
T_f  \simeq \alpha^2 \times \frac{M_m}{B^2}
\simeq 65.1~{\text {MeV}}.
\end{gather} 
This is below the muon decoupling temperature, which 
tells that the annihilation continues very long time. 

\begin{figure}
\includegraphics[height=4.5cm, width=7cm]{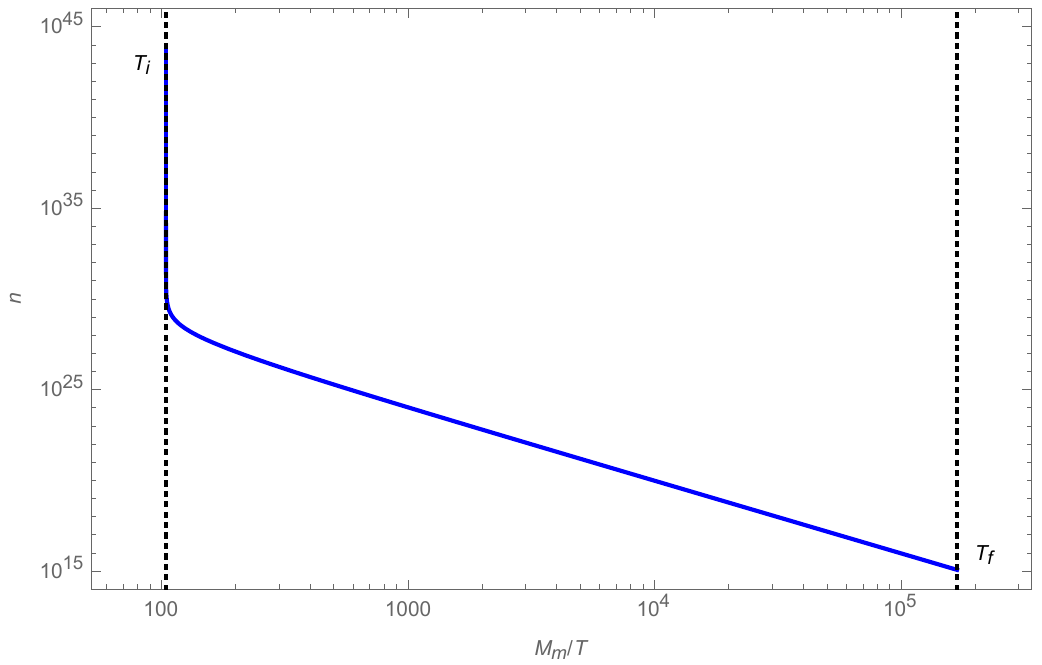}
\caption{\label{mden} The evolution of the monopole 
density $n$ (per $cm^{3}$) against $\tau=M_m/T$ with 
$M_m=11~\text{TeV}$.}
\end{figure}

From this we can estimate the remnant monopole density 
in the present universe. The monopole density after 
the annihilation around $T_f$ becomes
\begin{gather}
n_f \simeq \alpha^3 \times \frac{M_m}{BCM_P}~T_f^3 
\simeq 1.13 \times 10^{15}~cm^{-3}.
\label{fmden}
\end{gather}
where now we have put $M_m \simeq 11~{\rm TeV}$
since the temperature has cooled down very much. 
Obviously this is much lower than the initial 
monopole density given by (\ref{Cbound1}). 
The number of monopole within the co-moving volume 
is conserved thereafter. But they still interact 
with the electron pairs in the hot plasma before 
the decoupling around $T_d\simeq 0.5~\text{MeV}$, 
when the electron pairs disappear and the interaction rate becomes less than the Hubble expansion rate. 
The evolution of the monopole density $n$ against 
$M_m /T$ is shown in Fig. \ref{mden}, where we 
have put $M_m \simeq 11~{\rm TeV}$.  

Assuming that the expansion is adiabatic, the current 
number density and the energy density of the monopole 
is given by
\begin{gather}
n_0 = \frac{g_{s,0}}{g_{s,f}}
~\Big(\frac{T_0}{T_f}\Big)^3~n_f
\simeq \alpha^3~\frac{g_{s,0}}{g_{s,f}} 
\times \frac{M_m}{BCM_P}~T_0^3,
\label{mden0}
\end{gather}
where $g_s$ is the effective number of degrees of 
freedom in entropy and $T_0 = 2.73~\text{K}=2.35 
\times 10^{-13}~\text{GeV}$ is the temperature of 
the universe today. With $g_{s,f} \simeq 10.75$
and $g_{s,0} \simeq 3.9$, we have
\begin{gather}
n_0 \simeq 1.1 \times 10^{-23}~T_0^3
\simeq 1.9 \times 10^{-20}~ cm^{-3}.
\label{mden0}
\end{gather}
This means that there are roughly $6.6 \times 10^{66}$ monopoles in the observable universe (with the radius 
$4.4 \times 10^{28}~cm$), or roughly $2.04 \times 10^7$ monopoles per volume of the earth in the present universe. This implies that there are enough electroweak monopoles left over in the present universe that we could detect.
 
With (\ref{mden0}) the density parameter of monopole 
at present universe can be written as 
\begin{gather}
\rho_{m,0} =n_0~M_m 
\simeq 2.04 \times 10^{-7}~{\rm eV}~cm^{-3},  \nn\\
\Omega_m~h^2 =\frac{\rho_{m,0}}{\rho_\text{c,0}}~h^2
\simeq 4.3 \times 10^{-8},
\label{mden1}
\end{gather}
where $\rho_\text{c,0}={3H^2}/{8\pi G}
\simeq 1.05~h^2 \times 10^{-5}~{\rm GeV}~cm^{-3}
\simeq 0.48 \times 10^{-5}~{\rm GeV}~cm^{-3}$ is 
the critical density of present universe and 
$h \simeq 0.678$ is the scaled Hubble constant in 
the unit $H_0/(100~\text{km}~\text{s}^{-1}
~\text{Mpc}^{-1})$. This is about $0.38 \times 10^{-13}$ 
of the baryon number density $n_b \simeq 2.5 \times 10^{-7}~\text{cm}^{-3}$, which assures that the electroweak monopole cannot be a dark matter candidate. 

Actually, however, the free streaming remnant 
monopole density could be much less than the above 
estimate \cite{pta19}. There are good reasons for this. First, most of them could turn to the PMBHs 
and become the seed of the stellar objects and galaxies, as we will see in the following. So many 
of them might have been buried in stellar objects 
and galactic centers. The recent observations that there is a strong radial magnetic field near 
the galactic center could be the experimental evidences of this \cite{rmgc}. 

Second, the monopoles have a very short penetration 
length in the matter because they have strong 
magnetic interaction. In fact a relativistic electroweak monopole can travel only a few meters (less than 10 m) 
in Aluminium before they are trapped \cite{bolo}. This means that most of the monopoles left over which did not become the PMBHs could have been trapped and filtered 
out by the stellar objects, when they collide with them. This strongly implies that the actual free streaming remnant monopole density could be much less than 
(\ref{mden1}). In fact, the Parker bound on the free streaming monopole density implies that the remnant monopole density could be $10^{-4}$ times less than 
(\ref{mden1}) \cite{pta19,park}. Unfortunately it is difficult to estimate how much of them are free streaming 
at present universe.

Notice that, since the decoupling temperature of 
the electroweak monopole is much less than the monopole 
mass, the free streaming monopoles just after 
the decoupling start as completely non-relativistic. 
But eventually they are accelerated by the intergalactic magnetic field and become extremely relativistic.
So they can acquire the energy which exceeds 
the Greisin-Zatsepin-Kuzmin (GZK) energy limit, and 
thus could become identified as the ultra high energy 
cosmic rays \cite{plb24,ta,gzk}.

We can estimate the number of the monopoles which 
arrive on earth per year. Suppose the density of 
the free streaming monopoles at present universe 
per $km^3$ is $\bar n_0$. Let the radius of the earth 
be $r_0$ and consider a point located at the distance $r~(r>r_0)$ from the center of the earth. Since 
the solid angle of the earth viewed from this point 
is given by
\begin{gather}
\Omega (r) = 2\pi (r^2-r_0^2)\Big(1
-\frac{\sqrt{r^2-r_0^2}}{r} \Big),
\end{gather}
the number $n_{ES}$ of the monopoles coming from 
the sphere of radius $r_1~(r_1 > r_0)$ from the center 
of the earth to the earth surface is given by
\begin{gather}
n_{ES} = n_0 \Int_{r_0}^{r_1} r^2 
\frac{\Omega (r)}{4\pi (r^2-r_0^2)}
\sin \theta dr d\theta d \varphi \nn\\
= \frac{2\pi n_0}{3} 
~\Big[r_1^3-r_0^3 -(r_1^2-r_0^2)^{3/2}\Big]  \nn\\
= \frac{2}{3} 
~\Big[L^3+3L^2 r_0(1+\frac{r_0}{L}) 
-L^3(1+\frac{2r_0}{L})^{3/2} \Big]   \nn\\
\simeq \pi n_0 L r_0^2,~~~(r_0 << L),
\end{gather}
where we have put $L=r_1-r_0$. This means that 
the number of the monopoles arriving on earth per 
year (with $1~year \simeq 3.1558 \times 10^7~sec$ 
and $r_0\simeq 6400~km$) per $km^2$ is given by 
\begin{gather}
n_{ES}(1~year/km^2) \simeq n_0 \times 
\frac{3.1558 \times 3}{4 \times (6.4)^2} \times 10^8  \nn\\
\simeq 5.78~n_0 \times 10^6.
\end{gather}
So, with (\ref{mden0}) we have $n_{ES}/year~km^2
\simeq 10.75$. But we have to keep in mind that this estimate could be lower than this, since the density 
of the free streaming monopoles at present universe 
could be much lower than (\ref{mden0}). In this case 
the above estimate becomes consistent with the ultra 
high energy cosmic ray detection rate \cite{plb24,ta} 

Notice that the monopoles could create a huge density perturbation in the matter dominant era and thus become 
PMBHs \cite{pta19}. Indeed, with the monopole energy 
density $\rho_m$ and the energy density of the universe 
at the radiation-matter equality time $\rho_e$ given by 
\begin{gather}
\rho_m \simeq 11~{\rm TeV} \times \frac{3}{4\pi} M_W^3  
\simeq 1.78 \times 10^{50}~{\rm GeV}~cm^{-3}, \nn\\
\rho_e \simeq 5.9 \times 10^6~{\rm GeV}~cm^{-3},
\end{gather}
we have the monopole density perturbation
\begin{gather}
\frac{\delta \rho}{\rho} 
=\frac{\rho_m-\rho_e}{\rho_e} \simeq 0.3 \times 10^{44}.
\label{mdp}
\end{gather}
This strongly implies that the monopoles could turn to 
PMBHs \cite{pta19}.  

\section{Cho-Maison Monopole versus Primordial Reissner-Nordstrom Magnetic Blackhole}

Actually, for the Cho-Maison monopole to become a magnetic blackhole, we do not need the density perturbation. When coupled to gravity, it automatically turns to a magnetic blackhole \cite{bais,prd75,plb16,yangb,wong}. To show 
this, we start from (the bosonic sector of) 
the Einstein-Weinberg-Salam Lagrangian 
\begin{gather}
{\cal L}_{EWS} = \frac{\sqrt{-g}}{16\pi G} \Big\{R
-|{\cal D}_\mu \phi|^2-\frac{\lambda}{2}
\big(\phi^\dagger \phi-\frac{\mu^2}{\lambda} \big)^2  \nn\\
-\frac14 \F_\mn^2 -\frac{1}{4} G_\mn^2 \Big\}, \nn\\
{\cal D}_\mu \phi=\big(\pd_\mu
-i\frac{g}{2} \vec \tau \cdot \A_\mu 
- i\frac{g'}{2}B_\mu \big) \phi,
\label{ewslag}
\end{gather}
where $R$ is the scalar curvature, $\phi$ is the Higgs doublet, $\F_\mn$ and $G_\mn$ are the gauge field 
strengths of $SU(2)$ and $U(1)_Y$ with the potentials 
$\A_\mu$ and $B_\mu$, $g$ and $g'$ are the corresponding coupling constants. 

To proceed we choose the following spherically symmetric monopole ansatz in the spherical coordinates $(t,r,\theta,\varphi)$, 
\begin{gather}
\phi = \dfrac{i}{\sqrt{2}}\rho(r)~\left(\begin{array}{cc} 
\sin (\theta/2)~e^{-i\varphi} \\
- \cos(\theta/2) \end{array} \right),   \nn\\
\A_\mu= \frac{1}{g}(f(r)-1)~\hat r \times \pd_\mu \hr, \nn\\
B_\mu =-\frac{1}{g'}(1-\cos\theta) \pd_\mu \varphi.
\label{ans1}
\end{gather}
In terms of the physical fields the ansatz can be 
written as \cite{plb97,epjc15}
\begin{gather}
\rho=\rho(r),  \nn\\
A_\mu^{\rm (em)} = -\frac{1}{e}(1-\cos\theta) \pd_\mu \varphi,  \nn \\
W_\mu =\frac{i}{g} \frac{f(r)}{\sqrt2} e^{i\varphi}
(\pd_\mu \theta +i \sin\theta \pd_\mu \varphi), \nn\\
Z_\mu = 0.
\label{ans2}
\end{gather}
This clearly shows that the ansatz is for the electroweak 
monopole dressed by the W boson.

Now, adopting the static spherically symmetric space-time 
metric 
\begin{gather}
	ds^2 = -N^2 (r) A(r) dt^2 + \frac{dr^2}{A(r)} \nn\\
	+ r^2 (d^2\theta + \sin ^2 \theta d\varphi^2),  
\end{gather}
we find that the Einstein-Weinberg-Salam action reduces to  
\begin{gather}
	S=-\int N \Big[\frac{r\dot A+A-1}{8\pi G} 
	+ A K + U \Big] dr,	 \nn\\
	K = \frac{\dot{f}^2}{g^2 } 
	+  \frac{r^2}{2} \dot{\rho}^2,  \nn\\
	U = \frac{(1-f^2)^2}{2g^2 r^2} + \frac14 f^2 \rho^2
	+ \frac{\lambda}{8}r^2 (\rho^2 -\rho_0^2)^2   \nn\\
	+ \frac{1}{2g'^2 r^2}.
	\label{react}
\end{gather}
From this we have the following equations of motion 
\begin{gather}
	\frac{\dot{N}}{N} = 8\pi G \frac{K}{r}, \nn\\
	\dot A +\frac{A-1}{r} = -\frac{8 \pi G}{r} (A K + U), \nn\\
	A \ddot{\rho} + \Big(\dot{A} +\frac{\dot{N}}{N} A 
	+\frac{2A}{r} \Big) \dot{\rho} -\frac{f^2}{2r^2}\rho =\frac{\lambda}{2}(\rho^2-\rho^2_0)\rho,  \nn\\ 
	A \ddot{f} +\Big(\dot{A} +\frac{\dot{N}}{N} A \Big) \dot{f}
	+ \frac{1-f^2}{r^2 }f =\frac{1}{4}g^2 \rho^2 f.
	\label{gmeq}
\end{gather} 
Notice that, when the gravitation is switched off 
(i.e., when $G =0$ and $A=N=1$) this describes 
the non-gravitating Cho-Maison monopole which has 
the Coulombic monopole singularity at the origin \cite{epjc15}. 

When the gravity is included the above equation 
has two types of solutions, the globally defined gravitating Cho-Maison monopole solution and 
the RN type magnetic black hole solutions \cite{bais,prd75,pta19,plb16,yangb,wong,volk}. 
Consider the gravitating Cho-Maison monopole first. 
For this we may adopt the boundary condition
\begin{gather}
	A(0)=1,~~~f(0) =1,~~~\rho(0) =0,  \nn\\
	A(\infty) =1,~~~f(\infty) =0,~~~\rho(\infty) =\rho_0.
	\label{gcmbc}
\end{gather}
and obtain the following expansions for the solution 
near $r=0$,
\begin{gather}
	f(r) = 1- f_1 x^2+...,   \nn \\
	\rho(r) = h \rho_0 x^{\delta}+...,
	~~~\delta=\frac{\sqrt{3}-1}{2}, \nn \\
	A(r) =1-\frac{16\pi}{e^2} \sin^2 \theta_W 
	\Big(\frac{M_W}{M_P} \Big)^2 
	\frac{\delta^2}{\sqrt 3} h^2 x^{2\delta}+..., 
	\label{gcm}
\end{gather}
where $f_1$ and $h$ are constants, $x=M_W r$, and 
$M_P \simeq 1.22 \times 10^{19}~{\rm GeV}$ is the Planck 
mass. This tells that a gravitating Cho-Maison monopole solution which has the weak boson dressing for 
$0 < r < \infty$ exists. Notice, however, that the factor $(M_W/M_P)^2$ in $A$ is extremely small, which makes 
the metric almost flat. 

This tells that the gravitational modification is extremely small, almost non-existent. Only when the W boson mass 
$M_W$ becomes unrealistically large, of the order of 
the Planck mass, the gravitational modification becomes 
explicit here. From this we can conclude that 
the gravitating Cho-Maison monopole is almost identical 
to the non-gravitating Cho-Maison monopole. In particular, 
the gravity has practically no influence on the mass of 
the monopole. 

What is more relevant for us here is the second type, 
the RN type blackhole solutions carrying the magnetic 
charge $4\pi/e$ \cite{bais,prd75,pta19,plb16,yangb,wong}. 
To discuss this, we first swich off the weak bosons and 
let $f=0$ and $\rho=\rho_0$. With this we can easily 
solve (\ref{gmeq}) and find 
\begin{gather}
	N=1,  \nn\\
	A(r)=1-\frac{2GM}{r} +\frac{4\pi G}{e^2} \frac{1}{r^2}, 
\end{gather}
where $M$ is the ADM mass of the blackhole. This has 
the outer horizon 
\begin{gather}
	r_H =r_+ = MG + \sqrt{M^2 G^2-4 \pi G/e^2}, 
	\label{bhr}
\end{gather}
with
\begin{gather}
M \geq M_{eRN} =\frac{M_P}{\sqrt{\alpha}} 
\simeq 11.7 \times M_P \nn\\
\simeq 1.43 \times 10^{20}~{\rm GeV}
\simeq 2.55 \times 10^{-4}~g,
\end{gather}
where $M_{eRN}$ is the extremal (minimal) mass of 
the RN blackhole. From this we have
\begin{gather}
	M_{RN} = \frac{r_H}{2G} + \frac{2\pi}{e^2 r_H}, 
\end{gather}
so that the horizon of the extremal RN blackhole is 
fixed by  
\begin{gather}
	r_H^{eRN} =\frac{L_P}{\sqrt{\alpha}}  
	\simeq 1.87 \times 10^{-32} cm,
\end{gather}
where $L_P\simeq 1.6 \times 10^{-33} cm$ is the Planck 
length. 

This tells that the naked Cho-Maison monopole without 
the weak boson dressing, when coupled to gravity, becomes 
the RN blackhole which carries the magnetic charge $4\pi/e$, which can have any mass bounded below by the extremal mass $M_P/\sqrt{\alpha}$ \cite{bais,prd75}. So the gravity 
sets the minimum mass of the naked Cho-Maison monopole 
of the order of ten times the Planck mass, making it an extremal RN blackhole. 

Moreover, with non-trivial $f$ and $\rho$, the above 
RN monopole can be generalized to the modified RN 
blackhole which has the weak boson hair \cite{yangb,wong}. 
To see this we choose the boundary condition 
\begin{gather}
	A(r_H)=0,~~~f(r_H) =f_H,~~~\rho(r_H) =\rho_H,  \nn\\
	0 \leq f_H \leq 1,~~~~0 \leq \rho_H \leq \rho_0,   \nn\\
	A(\infty) =1,~~~f(\infty) =0,~~~\rho(\infty) =\rho_0.
	\label{mrnbc}	
\end{gather}
With this we have the following equation from (\ref{gmeq}),
\begin{gather}
	\dot A_H=\frac{1}{r_H} \Big(1-8\pi G U_H \Big),  \nn\\
	\dot A_H \dot \rho_H -\frac{f_H^2}{2r_H^2} \rho_H =\frac{\lambda}{2}(\rho_H^2-\rho^2_0)\rho_H,  \nn\\ 
	\dot A_H \dot f_H + \frac{1-f_H^2}{r_H^2 }f_H 
	=\frac{1}{4}g^2 \rho_H^2 f_H.
	\label{mrnc}	
\end{gather}
Now, assuming $\dot A_H \geq 0$, we have from the first equation 
\begin{gather}
	1 \geq 8\pi G \Big(\frac{\lambda}{8} \rho_0^4 r_H^2 
	+\frac{1}{2g'^2 r_H^2} \Big)  > \frac{4\pi G}{g'^2 r_H^2},
\end{gather} 
or
\begin{gather}
	r_H > \frac{\sqrt{4\pi G}}{g'} 
	=\cos~\theta_W \frac{L_P}{\sqrt{\alpha}}   
	\simeq 1.37 \times 10^{-32} cm.
\end{gather} 
This tells that $r_H$ is smaller than $r_H^{eRN}$ by 
the factor $\cos \theta_W$.

\begin{figure}
\includegraphics[width=8cm, height=6cm]{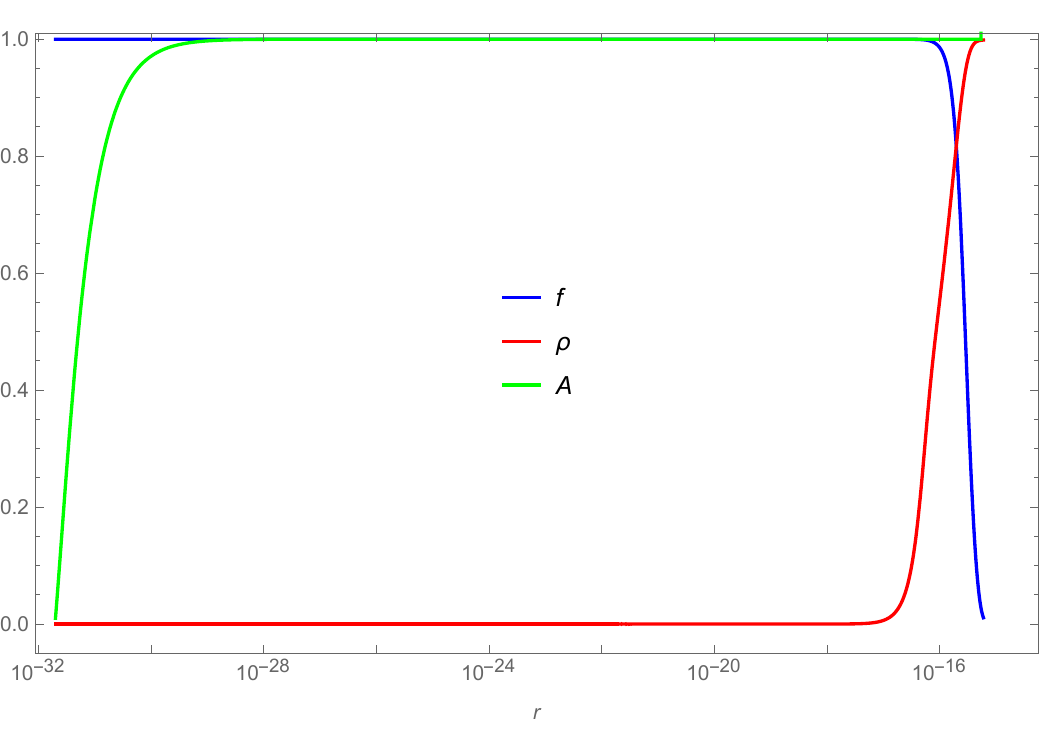}
\caption{\label{cmbh} The modified extremal Reissner-Nordstrom type magnetic blackhole in 
the Einstein-Weinberg-Salam theory which has the W boson $f$ (blue) and Higgs field $\rho$ (red) dressing.
Notice that with $N(r) \simeq 1$, $A(r)$ (green) 
represents the metric. The unit in the x-axis is 
$cm$.} 
\end{figure}

To proceed, we also assume that $f$ and $\rho$ are 
monotonically decreasing and increasing functions. 
With this we have the following constraint on $f_H$ 
and $\rho_H$ from the last two equations of (\ref{mrnc})
\begin{gather}
	\lambda \rho_0^2 r_H^2 \Big(1- \frac{\rho_H^2}{\rho_0^2} \Big)	
	< f_H^2 < 1-\frac{g^2}{4} \rho_H^2 r_H^2,  \nn\\
	\Big(\rho_0^2-\frac{f_H^2}{\lambda r_H^2} \Big) < \rho_H^2 
	< \frac{4}{g^2 r_H^2} (1-f_H^2)
\end{gather}
From this we have
\begin{gather}
	\Big(1 -\frac{g^2}{4\lambda} \Big)f_H^2 < 1 -\frac{g^2}{4} \rho_0^2 r_H^2.
\end{gather}
But the left-hand side is positive, since 
$g=e/ \sin \theta_W \simeq 0.65$ and $\lambda \simeq 0.26$
in the standard model. So we have
\begin{gather}
	r_H \leq \frac{1}{M_W} \simeq 2.5 \times 10^{-16}~cm.
\end{gather}
This means that the maximum horizon of the modified RN blackhole is set by the size of the Cho-Maison monopole. 
This is very important, because this tells that 
the blackhole is inside of the Cho-Maison monopole. 
Another outcome is that for the Cho-Maison monopole 
we have $f(0)=1$ and $\rho(0)=0$, but the boundary 
condition $f_H=1$ and $\rho_H=0$ is consistent with (\ref{gmeq}) only when $r_H \simeq 0.64/M_W$. 
The solution of the extremal RN blackhole whth the weak boson hair is shown in Fig. \ref{cmbh}.

To find the mass of the blackhole, notice that with 
(\ref{gmeq}) we have 
\begin{gather}
	A(r) =1- \frac{8\pi G}{r} \Big[e^{P(r)} 
	\int_{r_H}^r (K + U) e^{-P(r')}dr' + \frac{r_H}{2G} \Big],  \nn\\
	P(r) = 8\pi G \int_r^\infty \frac{K}{r'} dr',
	~~~(r_H \leq \ r),
\end{gather}
So, with (\ref{mrnbc}) the ADM mass of the modified 
RN blackhole is given by
\begin{gather}
	M_{mRN} =4\pi \int_{r_H}^\infty  (K + U) e^{-P(r)}dr
	+\frac{r_H}{2G},  
	\label{mint}
\end{gather}
which assures the positivity of the ADM mass. With 
this we can estimate the mass of the blackhole. When 
$r_H \ll 1/M_W$, we can switch off the gravity with 
flat metric, and calculate the leading contribution 
of the integral (\ref{mint}), and have
\begin{gather}
	M_{mRN} \simeq 0.75 \times \frac{4\pi \rho_0}{g} 
	+\frac{2\pi}{g'^2 r_H} +\frac{r_H}{2G}   \nn\\
	\geq \cos \theta_W \frac{M_P}{\sqrt {\alpha}} 
	+1.5 \times \sin^2 \theta_W \frac{M_W}{\alpha}  \nn\\
	\simeq 10.3~M_P +47.3~M_W 
	\simeq (1.3 \times 10^{17} +3.8)~{\rm TeV}  \nn\\
	\simeq 2.24 \times 10^{-4}~g. 
\end{gather}
This tells that the minimum mass of the modified Reissner-Nordstrom blackhole is set by the Planck mass, 
and that the contribution of the weak boson dressing 
to the mass is tiny (of $10^{-17}$), only 3.8 TeV. 
This is very interesting, because this 3.8 TeV is almost identical to the predicted Cho-Maison monopole mass 
3.96 TeV based on the scaling argument (without gravity) \cite{epjc15,pta19}.

On the other hand, when $r_H \simeq 1/M_W$, we have
\begin{gather}
	M_{mRN} \leq M_{mRN}^{max} 
	=\frac{M_P}{2M_W} M_P+ O(M_W)  \nn\\
	\simeq  9.3 \times 10^{32}~{\rm TeV} 
	\simeq 1.66 \times 10^{12}~g.	
\end{gather}
which sets the maximum mass of the blachkole. So the mass 
of the modified RN blackhole made of the Cho-Maison monopole can be anywhere between $10$ and $10^{36}$ times the Planck mass. It is interesting that the weak boson dressing with the gravity could set the range of the mass of the modified RN blackhole. Without the weak boson dressing the mass of the RN blackhole has only the lower bound. 

\section{Cosmic Production and Evolution of Electroweak Primordial Magnetic Blackhole}

Obviously the cosmic production and evolution of 
the electroweak monopoles do not apply to the above electroweak PMBHs. This is because the mass becomes 
totally different. As we have seen, the mass range 
of the modified RN type PMBH could be anywhere between 
$M_P/\sqrt \alpha$ (the extremal case) and
$(M_P/2M_W)~M_P$ (the non-extremal case). Moreover, 
the mass of the RN PMBH in principle has no upper 
limit. Now we discuss how this could change the initial density and evolution of the electroweak PMBHs.

Consider the extremal case first, whose mass and size 
is set by $1/\sqrt \alpha$ times the Planck scale. 
In this case the energy condition (\ref{econ}) should change to
\begin{gather}
E = \frac{\pi^2}{30}~g_*~T_i^4 \times \xi_{ebh}^3 
\geq \frac{M_P}{\sqrt \alpha}, 
\label{mbecon}
\end{gather} 
where $\xi_{ebh}$ is the size (the correlation length) 
of the extremal RN magnetic blackhole. From this we  
have 
\begin{gather}
\xi_{ebh} \simeq \Big(\frac{30}{\pi^2 g_* \sqrt \alpha} \frac{M_P^4}{T_i^4} \Big)^{1/3} \frac{1}{M_P} 
\simeq \frac{0.55 \times 10^{23}}{M_P} \nn\\
\simeq 0.89 \times 10^{-10}~cm 
\label{mbec1}
\end{gather} 
This is $0.8 \times 10^5$ bigger than the monopole correlation length $\bar \xi$ shown in (\ref{mcl1}) 
which determines the monopole density. Notice that 
for the non-extremal modified RN PMBH with mass 
$(M_P/2M_W)~M_P$, we have $\xi_{mbh} \geq 0.86 \times 10^{-4}~cm$. 

From this we have the initial number density $N_i$
of the extremal RN PMBH
\begin{gather}
	N_i^{ebh} \simeq \frac{g_P}{\xi_{ebh}^3}
	\simeq g_P \times 6.0 \times 10^{-69}~M_P^3  \nn\\
	\simeq 1.4 \times 10^{30} \times g_p~cm^{-3}.
	\label{ebhbound}
\end{gather}
This is smaller than the initial monopole density 
(\ref{Cbound1}) by the factor $1.8 \times 10^{-15}$. 
Similarly for the non-extremal modified RN PMBH 
with mass $(M_P/2M_W)~M_P$, we have $N_i^{mbh} 
\simeq 1.57 \times 10^{12} \times g_p~cm^{-3}$. 
This shows that there is a huge difference between 
the initial monopole density and magnetic blackhole 
density in the early universe. Again, this 
difference originates from the mass difference 
between the monopole and the magnetic blackhole.

It must be clear that these PMBHs are formed around 
the same time as the electroweak monopoles do, 
between $T_c$ and $T_G$, or around $T_i \simeq 102.09$
GeV. In time scale they are formed between 
$3.3 \times 10^{-11}~sec$ to $2.1 \times 10^{-10}~sec$ after the big bang for the period of 
$17.7 \times 10^{-11}~sec$, or around 
$6.7 \times 10^{-11}~sec$ after the big bang in average.

As we have pointed out, the mass of the RN PMBH theoretically has no upper limit, but obviously 
the number density of the PMBH depends on the mass. 
So one may ask what is the number density of RN PMBH 
with mass M. To answer this notice that in this case 
the energy condition (\ref{mbecon}) for the mass $M$ 
RN PMBH changes to    
\begin{gather}
E = \frac{\pi^2}{30}~g_*~T_i^4 
\times (\xi_i^{rnbh})^3 \geq M, 
\label{mbec2}
\end{gather} 
so that we have
\begin{gather}
\xi_i^{rnbh} \simeq \Big(\frac{30}{\pi^2 g_*} \frac{M^4}{T_i^4} \Big)^{1/3} \frac{1}{M}.
\label{mbec3}
\end{gather}
However, since $\xi_i^{rnbh}$ can not be bigger than 
the particle horizon in cosmology, we have
\begin{gather}
\Big(\frac{30}{\pi^2 g_*} 
\frac{M^4}{T_i^4} \Big)^{1/3} \frac{1}{M} 
\leq d_H(T_i)   \nn\\
\simeq 2~\Big(\frac{90}{32 \pi G g_*(T_i)} \Big)^{1/2}
~\frac{1}{T_i^2}.
\end{gather}
From this we have the maximum average mass of the RN 
PMBH,
\begin{gather}
M \leq M_{mmbh} = 0.5 \times 10^{32}~M_P \nn\\
\simeq 1.1 \times 10^{27}~g 
\simeq 0.55 \times 10^{-6}~M_\odot,
\label{mbhm}
\end{gather}
which is bigger than the maximum mass of the modified 
RN BH by the factor $10^{15}$. In this case we have
\begin{gather}	
r_H \simeq 1.0 \times 10^{32}~L_P \simeq 0.16~cm.
\label{mbh}
\end{gather}	
With this, the initial density of the RN PMBH of 
the maximum average mass $M_{mmbh}$ at $T_i$ is 
given by
\begin{gather}
N_i^{mmbh} \simeq \frac{g_P}{(\xi_i^{rnbh})^3}
\simeq \frac{g_P}{d_H(T_i)^3}  \nn\\
\simeq 1.3 \times 10^{-2} \times g_p~cm^{-3}.
\label{rnbhbound1}
\end{gather}
This should be compared with (\ref{ebhbound}).

It should be mentioned, however, that theoretically 
the maximum particle horizon for the RN PMBH mass is 
given (not by $d_H(T_i)$ but) by $d_H(T_G)$. This sets 
the upper limit of the RN PMBH mass $M$ in cosmology 
\begin{gather}
M \leq \bar M =1.57 \times 10^{32}~M_P \nn\\
\simeq 3.46 \times 10^{27}~g
\simeq 1.73 \times 10^{-6}~M_\odot.
\label{ulmbh}
\end{gather}	  
With this, the initial density $\bar N_i$ of the RN 
PMBH of the mass $\bar M$ at $T_i$ is given by
\begin{gather}
\bar N_i \simeq \frac{g_P}{(\xi_i^{rnbh})^3}
\simeq \frac{g_P}{d_H(T_G)^3}  \nn\\
\simeq 4.14 \times 10^{-4} \times g_p~cm^{-3}.
\label{ulbhbound}
\end{gather}
This should be compared with (\ref{rnbhbound1}). As we 
will see, this has deep implications in cosmology.

The PBH has become important in cosmology because 
there is a possibility that the PBH could account 
for the dark matter of the universe. It has been asserted that the PBH with the mass range 
$10^{17}~g$ to $10^{22}~g$ produced between $10^{-21}~sec$ and $10^{-16}~sec$ after the big 
bang (when the temperature becomes around $10^5$ 
to $10^7$ GeV) could account for all dark matter 
of the universe today \cite{carrk,green,carrc}. Moreover, recently it has been suggested that some 
of the PBH could carry net color charge, when 
the PBH is formed before the QCD confinement sets 
in above $T \simeq \Lambda_{QCD} \simeq 0.17$ 
GeV \cite{ak}. This has made the PBH more intersting object, and NASA is planning to hunt for such PBH 
with the new Roman Space Telescope \cite{pbh}. 
So one might wonder if our PMBH could also account 
for the dark matter, and could carry the color 
flux. 

We could think of two possible candidates for such 
PBHs in cosmology, the well known PBH proposed by 
Zeldivich and Novikov produced by the statistical 
density perturbation in the very early universe 
close to the Planck time \cite{zel,hawk,carr}, and 
our electroweak PMBH produced by the gravitational 
interaction of the electroweak monopole between 
$3.3 \times 10^{-11}~sec$ to $2.1 \times 10^{-10}~sec$ after the big bang. So the production mechanism 
and production period of two PBHs are totally different. And our PMBH carries the conserved magnetic charge, 
so that it can not evaporate. Moreover, with the mass 
range $2.55 \times 10^{-4}~g$ to $3.46 \times 10^{27}~g$, 
our electroweak PMBH could become a real candidate 
for the dark matter of the universe. To discuss
if our PMBH could really account for the dark matter, 
we have to discuss the cosmic evolution of the PMBH.    

The evolution of the PMBH is totally different from 
the evolution of the electroweak monopole. Here 
the annihilation of two magnetic blackholes which have opposite magnetic charges does not play any important 
role for two reasons. First, the correlation length 
of the blackhole $\xi_{bh}$ is roughly $10^5$ times 
bigger than that of the monopole, which means that 
the average distance between the monopole-antimonopole blackhole pairs is much bigger than the capture radius. Second, the mass of the magnetic blackholes is at least 
$10^{18}$ times bigger than the monopole. So the magnetic attraction between the blackhole pairs is not strong 
enough to make the annihilation. Moreover, even if 
they do, the two blackholes do not disappear. They 
merge to a magnetically neutral blackhole. This means 
that they evolve adiabatically with the Hubble expansion, 
with $N \propto 1/R^3$.   

A new aspect of the evolution of the PMBHs is 
the accretion and evaporation. Zeldovich and Novikov 
have suggested that their PBHs could have too much 
accretion from the nearby matters and become 
supermassive blackholes at present universe, which 
could be in conflict with the standard big bang 
cosmology \cite{zel,hawk,carr}. So we need to check 
if this could also appliy to our electroweak PMBH. 

Assuming that the accretion is a quasi-stationary 
process, one could express the rate of the accretion 
of the primordial blackhole during the radiation 
dominant era by \cite{zel,hawk,carr} 
\begin{gather}
\frac{dM}{dt}= \rho(t) \times 4\pi r_H^2
= \frac{2\pi^3}{15}~g_*(t)~T(t)^4 \times r_H^2  \nn\\
=\frac{3\pi^2}{8 G} \times \frac{r_H^2}{t^2}.	
\end{gather}
This, with (\ref{ardu}), (\ref{red}), and (\ref{bhr}), 
is written by
\begin{gather}
\frac{dx}{dt} =\frac{3\pi^2}{4 M_P} 
~\Big(1-\frac{2\pi}{e^2} \frac{1}{x^2}
+\sqrt{1-\frac{4\pi}{e^2} \frac{1}{x^2}} \Big) 
\times \frac{x^2}{t^2}   \nn\\
\simeq \frac{3\pi^2}{2 M_P} 
\times \frac{x^2}{t^2},	
\label{aeq}
\end{gather}
where $x=M/M_P$. Notice that the approximation in 
the second line follows from the fact that for our 
primordial blackhole $x$ is expected to be very large 
($x^2 \geq 1/\alpha$). Integrating this from the initial 
time $t_i \simeq 6.7 \times 10^{-11}~sec$ to 
the radiation-matter equilibrium time $t_e$ given 
by \cite{kolb}
\begin{gather}
t_e = 1.4 \times 10^3 (\Omega_0 h^2)^{-2}~years  
\simeq 2.1 \times 10^{11}~sec,
\end{gather}
we have the mass $M_e$ of the accreted blackhole at $t_e$ (with $t_i << t_e$),
\begin{gather}
M_e-M_i =\frac{3\pi^2}{2 M_P^2} 
\times \frac{M_e M_i}{t_e t_i} \times (t_e-t_i)  \nn\\
\simeq \frac{3\pi^2}{2 M_P^2} 
\times \frac{M_e M_i}{t_i}.
\label{eeb}
\end{gather}
So, for the extremal RN blackhole we have  
\begin{gather}
	M_e-M_i =\frac{3\pi^2}{2 \sqrt{\alpha}} 
	\frac{M_e}{M_P} \times \frac{1}{t_i}  
	\simeq 4.4 \times 10^{-38}~M_e,
\end{gather}
which means that the extremal blackhole has almost no accretion. Similarly for the modified RN blackhole with mass $(M_P/2M_W) M_P$ and horizon size $1/ M_W$ at 
$t_i$, the accretion becomes negligible. This is 
basically because the size of the extremal blackhole 
is very small compared to the particle horizon. 
Indeed, for the modified RN blackhole with horizon 
$1/M_W \simeq 2.5 \times 10^{-16}~cm$, the particle 
horizon $d_H$ at $t_i$ is given by $d_H(t_i) \simeq 4~cm$, bigger than the blackhole horizon by the factor $10^{16}$. 

For the maximum average mass of RN PMBH given 
by (\ref{mbhm}), however, we have
\begin{gather}
M_e-M_i \simeq \frac{3\pi^2}{2} 
\times \frac{M_i M_e}{M_P^2 t_i} \simeq 0.6~M_e,
\end{gather}
or
\begin{gather}
M_e \simeq 2.5~M_i \simeq 1.4 \times 10^{-6}~ M_\odot.
\end{gather}
In this case the PMBH does have a significant accretion, but does not become a supermassive blackhole. From this 
one might conclude that the electroweak PMBHs do not 
grow to a supermassive blackhole.   

Indeed, for (\ref{eeb}) to be valid, we must 
have
\begin{gather}
M_i < \frac{2 M_P t_i}{3\pi^2}~M_P
\simeq 0.84 \times 10^{32}~M_P  \nn\\
\simeq 1.83 \times 10^{27}~g.
\label{acc}
\end{gather}
Moreover, for the accretion to be effective, $M_i$ 
has to be very close to this value. Clearly this requirement is not easy to satisfy. This implies 
that our PMBHs are not likely to end up with 
supermassive blackholes at present universe. 

On the other hand it must be clear from (\ref{acc}) 
that, for our electroweak PNBH to become a supermassive blackhole, $M_i$ must be very close to $1.83 \times 10^{27}~g$. And the upper limit of the RN PMBH mass 
given by (\ref{ulmbh}) tells that some of our RN PMBH 
could meet with this condition. This is remarkable, 
because this tells that we can not completely exclude 
the possibility that our PMBHs could indeed grow 
to become supermassive blackholes.
 
Now, we consider the evaporation of the PMBHs. Hawking 
has argued that a PBH of mass $M$ has temperature 
of $10^{-6}~(M_\odot/M)~K$, so that blackholes with 
mass as large as $10^{15}~g$ would have radiated away 
all their mass by now \cite{hawk}. Fortunately this 
does not happen to our electroweak PMBHs, because 
they have the conserved magnetic charge which can not 
be evaporated. So the evaporation can only make them extremal.

This suggests that our electroweak PMBHs survive to 
the present universe in a wide range of blackholes, 
from the extremal RN type magnetic blackholes to supermassive magnetic blackholes. This strongly implies that the electroweak PMBHs could indeed become the seed 
of the stellar objects and galaxies, and could even 
account for the dark matter of the universe.   

\section{Remnant Electroweak Primordial Magnetic Blackhole}  

To test the plausibility that the electroweak PMBHs 
could account for the dark matter of the universe,
we have to know the density of the remnant electroweak PMBHs. As we have pointed out, the electroweak PMBHs (unlike the monopoles) have almost no annihilation 
after they are formed, so that the evolution equation 
of the PMBH becomes simple,
\begin{gather}
\frac{d N}{dt} + 3 H N = 0,
\label{boltz2}
\end{gather}
where $N$ is the number density of the PMBHs. In this 
case the number density of the PMBH at present universe 
is given by
\begin{gather}
N_0 = \frac{g_{s,0}}{g_{s,i}}
~\Big(\frac{T_0}{T_i}\Big)^3~N_i
\simeq 0.45 \times 10^{-45}~N_i
\label{bhden0}
\end{gather}
So, for the RN PMBMs with maximum average mass 
$M_{mmbh}=0.5 \times 10^{32}~M_P$ with the initial 
density given by (\ref{rnbhbound1}), we have 
$N_0 \simeq 0.59 \times 10^{-47}~g_P~cm^{-3}$. 
With this we have the energy density of the maximum 
average mass RN PMBH at $T_0$,
\begin{gather}
\rho_{mmbh} (T_0)=N_0 \times 0.5 \times 10^{32}~M_P \nn\\
\simeq 0.36 \times 10^{4}~{\rm GeV}~cm^{-3},  \nn\\
\Omega_{mmbh} =\frac{\rho_{bh,0}}{\rho_\text{c,0}}
\simeq 0.74 \times 10^{9}.
\label{edmmbh}
\end{gather}
This is too much, which implies that the PMBHs would overclose the universe. Certainly this is not 
acceptable.

A simple way to avoid this difficulty is to assume that 
the PMBHs have the Hawking radiation during the matter dominant era and reduce the mass. To test this idea, 
let us consider the maximum average mass PMBHs and 
suppose they (after the radiation dominant era) undergo 
the evaporation which reduces the mass to $M_{mmbh0}$. 
In this case the condition that the maximum average 
mass PMBHs not to overclose the universe is given by
\begin{gather}
\rho_{mmbh} (T_0)=N_0 \times M_{mmbh0} \leq \rho_{c,0} \nn\\
\simeq 0.48 \times 10^{-5}~{\rm GeV}~cm^{-3},
\end{gather}
or
\begin{gather}
M_{mmbh0} \leq 0.81 \times 10^{42}~g_P^{-1}~{\rm GeV} \nn\\
\simeq 0.67 \times 10^{23}~M_P.
\label{edmbh1}
\end{gather}
So, the maximum average mass PMBHs should evaporate 
and reduce the mass from $0.5 \times 10^{32}~M_P$ 
to less than $0.67 \times 10^{23}~M_P$, not to overclose 
the universe. This of course is less than the maximum 
average mass $M_{mmbh}$, but much bigger than the extremal blackhole mass $M_{ebh}=M_P/\sqrt \alpha$. Certainly 
this seems possible.

In fact, if we assume that the maximum average mass 
PMBHs become the extremal RN PMBHs after the evaporation, we have
\begin{gather}
\rho_{mmbh} (T_0)=N_0 \times \frac{M_P}{\sqrt \alpha}
\simeq 8.43 \times 10^{-28}~{\rm GeV}~cm^{-3},  \nn\\
\Omega_{mbh} =\frac{\rho_{bh,0}}{\rho_\text{c,0}}
\simeq 1.76 \times 10^{-22}.
\label{edmbh2}
\end{gather}
In this case the remnant PMBHs has almost no contribution to the energy density at present universe. This strongly implies that the evaporation of the PMBHs during 
the matter domonant era could allow them not to overclose the universe. This is nice.

Actually we could find the condition on the initial 
density of the PMBH at $T_i$ for the PMBHs not 
to overclose the universe. Suppose the mass of 
the PMBHs at present universe is $M_0$ in average. 
Then, from (\ref{bhden0}) we have 
\begin{gather}
\rho_{bh} (T_0)=N_0 \times M_0
= 0.45 \times 10^{-45}~N_i \times M_0 \nn\\
\leq \rho_{c,0} 
\simeq 0.48 \times 10^{-5}~{\rm GeV}~cm^{-3}.
\label{edmbhc}
\end{gather} 
With this we have
\begin{gather}
N_i \leq \frac{1.07 \times 10^{40}~{\rm GeV}}
{M_0}~cm^{-3}.
\label{edmbhc}
\end{gather}
So, assuming that the present PMBHs are extremal 
(i.e., assuming $M_0 =M_P/\sqrt \alpha$), we have 
\begin{gather}
N_i \leq 7.5 \times 10^{19}~cm^{-3}.
\label{edmbhc}
\end{gather}  
From this we can deduce
\begin{gather}
\xi_i^{ebh} \geq  0.24~g_P^{1/3} \times 10^{-6}~cm,
\end{gather}  
which implies that the initial mass $M_i^{ebh}$ at $T_i$
could be around
\begin{gather}
M_i^{ebh} \geq \frac{\pi^2}{30}~g_*~T_i^4 
\times (\xi_i^{ebh})^3 
\simeq 6.57 \times 10^{30}~{\rm GeV}  \nn\\
\simeq 5.39 \times 10^{21}~M_P.
\end{gather} 
Unfortunately, at the moment we have no way to predict 
the average mass $M_0$ of the remnant PMBHs at present universe.

The above discussion implies that $\Omega_{mbh}$ of 
the PMBHs could have the value similar to that of 
the dark matter, and could account for the dark matter 
of the universe. And this could happen within 
the framework of the standard cosmology. This is 
a very interesting possibility. But, of course, 
this is a proposal (a possibility) which has yet 
to be verified. Ideally, we should be able to predict 
the initial and final density of the electroweak 
PMBHs, and predict the average mass of the PMBHs. 
Unfortunately we could not do this in this work. 
Instead, we provided the plausibility argument 
how this could happen with the electroweak PMBHs. 
What is  comforting is that there seems no obvious 
obstacle which can forbid this.  

\section{Physical Implications}

In this paper we have compared the cosmic production 
of the electroweak monopoles and the electroweak 
PMBHs, and discussed the cosmological implications of 
the electroweak PMBHs. In particular, we have studied 
the possibility that the electroweak PMBHs become 
the seed of the stellar objects and galaxies, and 
account for the dark matter of the universe. 
The justification of this study is that, if the standard 
model is correct, we can not avoid the electroweak 
PMBH in cosmology. This is because the electroweak 
monopole automatically turns to the RN type PMBH 
when coupled to the gravity. 

The cosmological implications of the electroweak 
monopole has been studied before, and it has been 
suggested that the electroweak monopoles could 
generate the density perturbation and turns to
the PMBHs, and thus become the seed of the stellar 
objects and galaxies \cite{pta19}. In this paper 
we have re-analized the cosmic production of 
the electroweak monopole in the early universe
to argue that the popular monopole production mechanisms, 
the vacuum bubble collisions in the first order phase transition and the Kibble-Zurek mechanism in the second order phase transition, may have serious defects. 
This is because the monopoles are created by the thermal fluctuations of the Higgs vacuum which could continue 
long after the phase transition till the temperature 
cools down to the Ginzburg temperature. 

This suggests that in both first and second order 
phase transitions the major monopole production could 
take place after the phase transition. In this case  
the main monopole production in the first order phase 
transition could take place after the phase 
transition after the vacuum tunneling, as far as 
the Ginzburg temperature is considerably lower than 
the critical temperature. This means that what is 
important in the monopole production in cosmology 
is not the type of the phase transition but what is 
the Ginzburg temperature. 

We have also argued that another popular wisdom, 
the assumption that the initial monopole density is 
determined by the correlation length fixed by the Higgs 
mass, may have a shortcoming. This is because the mass 
of the monopole is given by $1/\alpha$ times 
the W boson mass but the size of the monopole is fixed 
by the W boson mass, so that the energy within 
this correlation volume may not be enough to create 
the monopole. To cure this defect we have suggested 
to adopt a new correlation length $1/\alpha^{1/3}$ 
times bigger than the correlation length fixed by 
the Higgs mass. 

Our study tells that the cosmic evolution of 
the electroweak PMBH is totally different from that 
of the electroweak monopole. There are many reasons for this. First, the electroweak PMBH has a huge mass with a wide range of uncertainty at the initial stage, from  $M_P/\sqrt \alpha \simeq 2.55 \times 10^{-4}~g$ (for the extremal PMBH) to $1.57 \times 10^{32}~M_P \simeq 3.46 \times 10^{27}~g$, compared to 
the monopole mass fixed by 11 TeV. This changes 
the initial density of the PMBHs greatly. 
Second, unlike the monopoles, the PMBHs have 
virtually no annihilation at the initial stage, basically because the monopole-antimonopole capture radius becomes much bigger than the correlation 
length which determines the initial monopole density. 

Instead, what is important for the PMBHs is 
the accretion and evaporation. We have shown that 
our PMBHs could have significant accretion which 
in principle could make some of PMBHs super massive 
blackholes under certain conditions. As for the evaporation, the electroweak PMBHs can not evaporate completely, 
because they carry the conserved magnetic charge. 
So the evaporation could only make them the extremal 
RN blackhole. On the other hand the evaporation could 
play important roles in cosmology. All in all, we have 
argued that the electroweak PMBHs could become the seed 
of the stellar objects and galaxies, and could even 
account for the dark matter of the universe without conflicting with the standard cosmology.

Unfortunately, we have provided only the plausibility 
argument how this could happen in this paper. One 
reason why we could not prove this is the huge 
uncertainty in the mass range of the PMBH. Another 
reason is that, in principle we have two possible 
solutions, the PMBH and the gravitationally modified 
electroweak monopole which is almost identical 
to the electroweak monopole, when the electroweak 
monopole is coupled to gravity. So we do not know 
how many electroweak PMBHs we could have at the initial 
stage of the monopole formation in cosmology. These 
are the big uncertainties that we have to remove 
before we could make a precise prediction.    

But a most important point of our study is that in 
cosmology we have a new type of PBH, the electroweak PMBH, which is totally different from the popular 
PBH proposed by Zeldovich and Novikov. This PBH is 
a prediction based on the density perturbation, 
but our PMBH is an inevitable outcome of the standard 
model. So, if the standard model is correct, we have 
to deal with this. Moreover, our study in this paper  tells that the mass range of our PMBH is from 
$2.55 \times 10^{-4}~g$ to $3.42 \times 10^{27}~g$, which covers the mass range $10^{17}~g$ to 
$10^{22}~g$ of the popular PBH dark matter 
candidate \cite{carrk,green,carrc}. This strongly suggests that our electroweak PMBHs could become 
ideal featherweight mini PBH which could actually account for the dark matter, and could even carry 
the color flux. 

This brings us interesting questions. Do we really need different types of primordial blackholes in cosmology? Which could be more realistic? Could 
they transform to each other merging together or absorbing the other? Can our PMBH could also carry 
the color charge? If so, how? And we can ask more questions. Certainly we need more research work to answer these questions.  
 
{\bf ACKNOWLEDGEMENT}

~~~This work is supported in part by the National Research Foundation of Korea funded by the Ministry of Science and Technology (Grant 2022-R1A2C1006999), Center for Quantum Spacetime, Sogang University, and Department of Liberal Arts, Korea Polytechnic University, Korea.


\begin{thebibliography}{99}
\bibitem{dirac} P.A.M. Dirac, Proc. Roy. Soc. London, 
{\bf A133}, 60 (1931); Phys. Rev. {\bf 74}, 817 (1948).
\bibitem{cab} B. Cabrera, Phys.  Rev. Lett.  {\bf 48}, 
1378 (1982). 

\bibitem{wu} T.T. Wu and C.N. Yang, in {\it Properties 
of Matter under Unusual Conditions}, edited by 
H. Mark and S. Fernbach (Interscience, New York) 1969; 
Phys. Rev. {\bf D12}, 3845 (1975);
Y.M. Cho, Phys. Rev. Lett. {\bf 44}, 1115 (1980); 
Phys. Lett. {\bf B115}, 125 (1982).
\bibitem{thooft} G. 't Hooft, Nucl. Phys. {\bf B79}, 276 (1974); A.M. Polyakov, JETP Lett. {\bf 20}, 194 (1974); 
B. Julia and A. Zee, Phys. Rev. {\bf D11}, 2227 (1975);
M. Prasad and C. Sommerfield, Phys. Rev. Lett. {\bf 35}, 
760 (1975).
\bibitem{dokos} C. Dokos and T. Tomaras, Phys. Rev. 
{\bf D21}, 2940 (1980).
\bibitem{plb97} Y.M. Cho and D. Maison, Phys. Lett. 
{\bf B391}, 360 (1997).
\bibitem{yang} Yisong Yang, Proc. Roy. Soc. London, 
{\bf A454}, 155 (1998); Yisong Yang, {\it Solitons 
in Field Theory and Nonlinear Analysis} (Springer 
Monographs in Mathematics), p. 322 (Springer-Verlag) 2001.

\bibitem{epjc15} Kyoungtae Kimm, J.H. Yoon, and Y.M. Cho, 
Eur. Phys. J. {\bf C75}, 67 (2015).

\bibitem{ellis} J. Ellis, N.E. Mavromatos, and T. You, 
Phys. Lett {\bf B756}, 29, (2016).
\bibitem{bb} F. Blaschke and P. Benes, Prog. Theor. 
Exp. Phys. 073B03 (2018).
\bibitem{epjc20} Pengming Zhang, Liping Zou, and Y.M. Cho, Euro. Phys. J. {\bf C80}, 280 (2020).
\bibitem{bbc} P. Benes, F. Blaschke, and Y.M. Cho, 
arXiv:2024.10747[hep-ph], to be published.

\bibitem{medal1} B. Acharya et al. (MoEDAL Collaboration),
Phys. Rev. Lett. {\bf118}, 061801 (2017).
\bibitem{medal2} B. Acharya et al. (MoEDAL Collaboration),
Phys. Rev. Lett. {\bf 123}, 021802 (2019). 
\bibitem{medal3} B. Acharya et al. (MoEDAL Collaboration), Nautre {\bf 602}, 63 (2022).
\bibitem{atlas} G. Aad. et al. (ATLAS Collaboration), 
Phys. Rev. Lett. {\bf 124}, 031802 (2020). 

\bibitem{kibb} T.W.B. Kibble, J. Phys. {\bf A9} 1387 (1976).
\bibitem{pres} J.P. Preskill, Phys. Rev. Lett. {\bf 43}, 1365 (1979). 
\bibitem{guth} A.H. Guth and E.J. Weinberg, Nucl. Phys. 
{\bf B212}, 321 (1983).
\bibitem{zurek}  W.H. Zurek, Phys. Rep. {\bf 276} 177 (1996).
\bibitem{infl} A.H. Guth, Phys. Rev. {\bf D23}, 347 (1981);
A.D. Linde, Phys. Lett. {\bf B108} 389 (1982).

\bibitem{pta19} Y.M. Cho, Phil. Trans. R. Soc. {\bf A377}, 0038 (2019).
\bibitem{plb24} Y.M. Cho and Franklin H. Cho, 
Phys. Lett {\bf B851}, 138598, (2024).

\bibitem{bais}F. Bais and R. Russel, Phys. Rev. {\bf D11}, 2692 (1975).  
\bibitem{prd75} Y.M. Cho and P.G.O. Freund, Phys. Rev. {\bf D12}, 1588 (1975).
\bibitem{plb16} Y.M. Cho, Kyoungtae Kimm, J.H. Yoon, Phys. 
Lett. {\bf B761}, 203 (2016).
\bibitem{yangb} Yang Bai and Mrunal Korwar, JHEP {\bf 04},
119 (2021).
\bibitem{wong} D. Zhu, K.M. Wong and G.Q. Wong, Commun. Theor. Phys. {\bf 76}, 035201 (2024).
\bibitem{volk} R. Gervalle and M. Volkov, Phys. Rev. Lett. {\bf 133},191402 (2024).
%\bibitem{bfm}
\bibitem{zel} Y.B. Zeldovich and I.D. Novikov,
Sov. Astro. AJ {\bf 10}, 602 (1967).
\bibitem{hawk} S. Hawking, Nature {\bf 248}, 30 (1974).
\bibitem{carr} B. Carr and S. Hawking, Mon. Not. R. Astr. Soc. {\bf 168}, 399 (1974).
\bibitem{carrk} B. Carr and F. Kuhnel, Annu. Rev. Nucl. Part. Sci. {\bf 70}, 355 (2020).
\bibitem{green} A.M. Green and B.J. Kavanagh,
J. Phys. {\bf G48}, 043001 (2021).
\bibitem{carrc} B. Carr, S. Clesse, J. Garcia-Bellido,
M. Hwakins, and F. Kuhnel, Phys. Rep. {\bf 1054}, 1 (2024).

\bibitem{ak} E. Alonso-Monsalve and D. Kaiser,
Phys. Rev. Lett. {\bf 132}, 230402 (2024).
\bibitem{pbh} W. DeRocco, E. Frangipane, N. Hamer,
S. Profumo, N. Smyth, Phys. Rev. {\bf D109}, 023013 (2024).

\bibitem{kriz} D.A. Krizhnits and A.D. Linde, Phys. Lett. 
{\bf B42}, 471 (1972); C. Bernard, Phys. Rev. {\bf D9}, 3312 (1974).
\bibitem{and} G. Anderson and L. Hall, Phys. Rev. 
{\bf D45}, 2685 (1992); M. Dine, R. Leigh, P Huet, 
A. Linde, and D. Linde, Phys. Rev. {\bf D46}, 550 (1992).
\bibitem{koba} S. Arunasalam and A. Kobahhidze,
Euro. Phys. J. {\bf C77}, 444 (2017); S. Arunasalam, 
D. Collision, and A. Kobahhidze, hep-ph/1810.10696 
(2018).

\bibitem{gin} V. Ginzburg, Sov. Phys. Solid State {\bf 2}, 1824 (1960).
\bibitem{kolb} See, e.g., E. Kolb and M. Turner, 
{\it The Early Universe} (Addition-Wesley Publishing Co.) 1990.

\bibitem{vil} A. Vilenkin and E. Shellard,  
{\it Cosmic Strings and other Topological Defects} 
(Cambridge University Press) 1994.

\bibitem{rmgc} R. Eatough et al., Nature {\bf 591}, 
391 (2013); M. Zamaninasab, E. Clausen-Brown, 
T. Savolainen, A. Tchekhovskoy, Nature {\bf 510}, 126 (2014).
\bibitem{bolo} S. Cecchini, L. Patrizii, Z. Sahnoun, G. Sirri, and V. Togo, arXiv: hep-ph/1606.01220 (2016).
\bibitem{park} E.N. Parker, Astrophys. J. 160, 383 (1970).

\bibitem{ta} Telescope Array Collaboraltion,
Science {\bf 382}, 903 (2023).

\bibitem{gzk} K. Greisen, Phys. Rev. Lett. {\bf 16}, 
748 (1996); G. Zatsepin, V. Kuzmin, Pisma Zh. Eksp. 
Teor. Fiz. {\bf 4}, 114 (1996).

\end{thebibliography}
\end{document}